\documentclass[preprint,12pt]{elsarticle}
\usepackage{amssymb,lineno,hyperref}
\usepackage[T1]{fontenc}
\graphicspath{{figs/}}
\begin{document}
\begin{frontmatter}
\title{Polysiloxane-based scintillators for shashlik
  calorimeters}

\author[FBK]{F.~Acerbi}
\author[INFNPD,uniPD]{A.~Branca}
\author[INFNBicocca,uniBicocca]{C.~Brizzolari}
\author[INFNPD]{G.~Brunetti}
\author[uniPD,INFNLegnaro]{S.~Carturan}
\author[INFNBari]{M.G.~Catanesi}
\author[INFNBo]{S.~Cecchini}
\author[INFNBo]{F.~Cindolo}
\author[INFNPD,uniPD]{G.~Collazuol}
\author[INFNPD]{F.~Dal~Corso}
\author[INFNNA,uniNA]{G.~De~Rosa}
\author[INFNPD,uniPD]{C.~Delogu}
\author[INFNBicocca,uniBicocca]{A.~Falcone}
\author[FBK]{A.~Gola}
\author[CENBG]{C.~Jollet}
\author[Zagabria]{B. Kli\v{c}ek}
\author[INR]{Y.~Kudenko}
\author[INFNPD,uniPD]{M.~Laveder}

\author[INFNPD,uniPD]{A.~Longhin\corref{corr}} \ead{andrea.longhin@pd.infn.it}

\author[INFNRM]{L.~Ludovici}
\author[INFNBicocca,Insubria]{E.~Lutsenko}
\author[INFNBari,uniBari]{L.~Magaletti}
\author[INFNBo]{G.~Mandrioli}
\author[INFNLegnaro]{T.~Marchi}
\author[INFNBo]{A.~Margotti}
\author[INFNBicocca,Insubria]{V.~Mascagna}
\author[CENBG]{A.~Meregaglia}
\author[INFNPD]{M.~Mezzetto}
\author[CERN]{M.~Nessi}
\author[INFNPD,uniPD]{M.~Pari}
\author[INFNBicocca,uniBicocca]{E.~Parozzi}
\author[INFNBo,uniBo]{L.~Pasqualini}
\author[FBK]{G.~Paternoster}
\author[INFNBo]{L.~Patrizii}
\author[FBK]{C.~Piemonte}
\author[INFNBo]{M.~Pozzato}
\author[INFNBicocca,Insubria]{M.~Prest}

\author[INFNPD]{F.~Pupilli\corref{corr}}
\ead{fabio.pupilli@pd.infn.it}

\author[INFNBari]{E.~Radicioni}
\author[INFNNA,uniNA]{C.~Riccio}
\author[INFNNA,uniNA]{A.C.~Ruggeri}
\author[INFNPD,uniPD]{C.~Scian}
\author[INFNBo]{G.~Sirri}
\author[Zagabria]{M.~Stip\v{c}evic}
\author[INFNBo]{M.~Tenti}
\author[INFNBicocca,uniBicocca]{F.~Terranova}
\author[INFNBicocca,uniBicocca]{M.~Torti}
\author[INFNBicocca]{E.~Vallazza}
\author[INFNPD,uniPD]{M.~Vesco}

\address[FBK]{Fondazione Bruno Kessler (FBK) and INFN TIFPA, Trento, IT}
\address[INFNPD]{INFN Sezione di Padova, via Marzolo 8, Padova, IT}
\address[uniPD]{Phys. Dep. Univ. di Padova, via Marzolo 8, Padova, IT}
\address[INFNBicocca]{INFN, Sezione di Milano-Bicocca, Piazza della Scienza 3, Milano, IT}
\address[Insubria]{DISAT, Univ. degli Studi dell'Insubria, via Valeggio 11, Como, IT}
\address[INFNLegnaro]{INFN, Laboratori Nazionali di Legnaro, Viale dell'Universit\`{a} 2, Legnaro (PD), IT}
\address[INFNBari]{INFN Sezione di Bari, via Amendola 173, Bari, IT}
\address[INFNBo]{INFN, Sezione di Bologna, viale Berti-Pichat 6/2, Bologna, IT}
\address[INFNNA]{INFN, Sezione di Napoli, Via Cintia, Napoli, IT}
\address[uniNA]{Phys. Dep., Univ. ``Federico II'' di Napoli, Napoli, IT}
\address[uniBicocca]{Phys. Dep. Univ. di Milano-Bicocca, Piazza della scienza 3, Milano, IT}
\address[CENBG]{CENBG, Universit\`e de Bordeaux, CNRS/IN2P3, 33175 Gradignan, FR}
\address[Zagabria]{Center of Excellence for Advanced Materials and Sensing Devices, Ru\dj er Bo\v{s}kovi\'{c} Institute, HR-10000 Zagreb, HR}
\address[INR]{Institute of Nuclear Research of the Russian Academy of Science, Moscow, RU}
\address[INFNRM]{INFN, Sezione di Roma 1, piazzale A. Moro 2, Rome, IT}
\address[uniBari]{Phys. Dep. Univ. di Bari, via Amendola 173, Bari, IT}
\address[CERN]{CERN, Geneva, Switzerland}
\address[uniBo]{Phys. Dep. Univ. di Bologna, viale Berti-Pichat 6/2, Bologna, IT}

\cortext[corr]{Corresponding authors}

\begin{abstract}
  We present the first application of polysiloxane-based scintillators
  as active medium in a shashlik sampling calorimeter. These results were obtained from a testbeam campaign of a $\sim 6 \times 6\times
  45$~cm$^3$ (13~$X_0$ depth) prototype.  A Wavelength Shifting
  fiber array of 36 elements runs perpendicularly to the stack of iron
  (15~mm) and polysiloxane scintillator (15~mm) tiles with a density
  of about one over cm$^2$.  Unlike shashlik calorimeters based on
  plastic organic scintillators, here fibers are optically matched
  with the scintillator without any intermediate air gap.  The prototype features a
  compact light readout based on Silicon Photo-Multipliers embedded in
  the bulk of the detector. The detector was tested with electrons, pions and
  muons with energies ranging from 1 to 7~GeV at the CERN-PS. This solution offers a highly radiation hard detector to instrument the decay region of a neutrino beam, providing an event-by-event measurement of
  high-angle decay products associated with neutrino production
  (ENUBET, Enhanced NeUtrino BEams from kaon Tagging, ERC project).
  The results in terms of light yield, uniformity and energy
  resolution, are compared to a similar calorimeter built with
  ordinary plastic scintillators.
\end{abstract}

\end{frontmatter}

%\linenumbers

\section{Introduction}
\label{intro}

Shashlik calorimeters~\cite{ref1, ref2} have been used since more than
20 years in particle physics~\cite{ref3,ref4,ref5,ref6,ref7,ref8}. These devices are sampling calorimeters in which the scintillation light is readout by WaveLength Shifting (WLS) fibers running perpendicularly to the stack of absorber and scintillator and hosted in holes through these elements.

These calorimeters have been considered in the context of
the ENUBET~\cite{ref30,enubet,enubet_eoi,enubet_loi_2018} project, where they could be employed to monitor positron production in the decay tunnel of conventional neutrino beams. Thanks to their robustness and good performance/cost ratio they could be effectively used over large surfaces to perform a precise measurement of the $\nu_e$ flux originating from kaon decays ($K^+ \to
e^+\pi^0\nu_e$)~\cite{ref30}. For this application, a longitudinal
segmentation of about 4~$X_0$ is needed to separate positrons from
charged pions with a misidentification probability
below~3\%~\cite{ref21}.  The most cost-effective solution exploits a
transverse modularity of 3$\times$3~cm$^2$ tiles. Since 2016, the ENUBET Collaboration has carried on an extensive experimental
campaign of tests at the CERN-PS beamlines employing prototypes with
standard plastic organic scintillators. In this context, the possibility of employing polysiloxane based scintillators instead of plastic scintillators is extremely appealing. This choice would ease the fabrication of the
scintillators, allow a perfect optical match between the fibers and
the scintillators, and ensure enhanced radiation hardness for the ENUBET application, where doses up to fractions of kGy are expected.

After a brief overview on the state of the art on
polysiloxane scintillators (Sec.\ref{PSsec}) and compact shashlik
calorimeters with standard plastic scintillator (Sec.\ref{SHsec}), we
make a comparison between the light yield of polysiloxane and plastic
scintillator using an Am source (Sec.~\ref{sources}). In
Sec.~\ref{optisim} we present the results of a simulation on light
collection efficiency for polysiloxane or plastic shashlik
configurations. In Sec.~\ref{setup} we describe the construction of
the polysiloxane-based shashlik calorimeter (\ref{POLY}) and of a
reference traditional shashlik calorimeter based on plastic
scintillators~(\ref{setupref}). The results of the particle exposure
at the CERN-PS T9 beamline are described in Secs.~\ref{t9} and
~\ref{results}.  Conclusions and future opportunities are summarized in
Sec.~\ref{conclusions}.

\section{Polysiloxane scintillators}
\label{PSsec}
Polysiloxane polymers (also known as silicones) are composed by a main
backbone, formed by Si and O atoms, regularly alternated, and by
organic substituents, attached to the silicon atoms of the main chain.
Polysiloxane mechanical properties are typical of elastomeric
materials and arise from the high degree of flexibility of the Si-O-Si
bridge, thus siloxanes have enhanced deformability and elongation at
rupture~\cite{Tiwari2014}.

On the other hand, this bond has enhanced strength with respect to C-O
and C-C bond and this partially accounts for remarkably high radiation
resistance~\cite{Bowen1989a,quarl} as compared to common organic polymers traditionally used
for plastic scintillators (i.e. PolyVinylToluene, PolyStyrene).  In standard plastic scintillators, transparency losses (yellowing) are
due to the production of free radicals from the breaking of C-H
or C-C bonds by irradiation and further recombination to produce
double bonds, acting as absorbing centers. The stronger Si-O bond
preserves transparency up to much higher doses. Radiation hardness is further increased by the presence of phenyl side groups absorbing part of the radiation energy thanks to the $\pi$ electrons in the aromatic
ring, thus reducing the damages along the polymeric
chain~\cite{Bowen1989a, Quaranta2013, Quaranta2010a}.

Suitable dyes to achieve scintillation light from incoming particles
are dissolved in the precursor siloxane resin, then the cross-linking
reaction is carried out using Pt-based catalyst~\cite{Carturan2011}.
Thanks to their viscous nature polysiloxane scintillators can be
simply poured in liquid tight containers and allowed to cross-link at
moderate temperatures finally leading to a solid, though soft and
rubbery, material.

\subsection{Previous R\&D on polysiloxane scintillators}

The development of polysiloxane scintillators began by Bowen {\it et
  al.}~\cite{Bowen1989a,Bowen1989b}. They report no yellowing with
doses up to 100~kGy and light outputs ranging from less than 5\% to
90\% of a commercial reference sample (BC-408).
The same group performed an extensive series of studies on the
radiation hardness of polysiloxane scintillators doped with different
types of primary dye,~\cite{Feygelman1990a, Walker1989}, secondary
dyes~\cite{Feygelman1990b} and different phenyl
concentrations~\cite{Harmon1991}.
More than ten years later, Bell {\it et al.}  exploited polysiloxane
scintillators for thermal neutron detection, through the loading with
B or Gd~\cite{Bell2002,Bell2004}.
More recently a detailed analysis was performed focusing on the
optimization of the scintillation performances, by varying the
composition and concentration of the mixture and
matrix~\cite{Quaranta2013,Quaranta2010a,Quaranta2010b};
on the performances for thermal neutron detection with the addition of
o-carboranes at different concentrations~\cite{Quaranta2011} and
$^{6}$LiF nanoparticles~\cite{CarturanNIMA19} and for fast neutron
detection with the Time-Of-Flight technique~\cite{Carturan2011}. Very
recently, the possibility to discriminate fast neutrons from $\gamma$s
on the basis of different light pulse shapes has been
demonstrated~\cite{tmarchi}.

\section{Shashlik calorimeters with embedded light readout}
\label{SHsec}
Recent developments in the technology of silicon-based photosensors
have allowed new solutions for light collection and
readout~\cite{ref9,ref12,ref13,ref14} and a broader range of applications is at
hand for shashlik detectors~\cite{enubet_eoi,ref15}.  The INFN SCENTT
Collaboration has developed an ultra-compact module (UCM) where every
single fiber segment is directly connected to a Silicon
PhotoMultiplier (SiPM) thus avoiding dead regions due to fiber
bundling with a net improvement in the homogeneity of the longitudinal
sampling. An extensive experimental campaign 
(\cite{ref18,ref20,papernov2016,paperIrrad}) at the CERN-PS has
followed within the ENUBET Collaboration.  The UCM concept has also been adopted
for the polysiloxane based shashlik calorimeter described in this
work.

\section{Light yield of polysiloxane with radioactive sources}
\label{sources}
The relative light yields of polysiloxane and standard plastic
scintillators have been characterized in the past~\cite{Carturan2011} using
$\alpha$ emitters. The measured light yield was found to be typically
lower in Polysiloxane scintillators by a factor of about two with respect
to Eljen plastic scintillators EJ-212 or EJ-200 (Eljen~\cite{ref22}).

\begin{figure}%[hbpt!]
\begin{minipage}{7cm}
\centering
$\vcenter{\includegraphics[scale=0.25,type=png,ext=.png,read=.png]{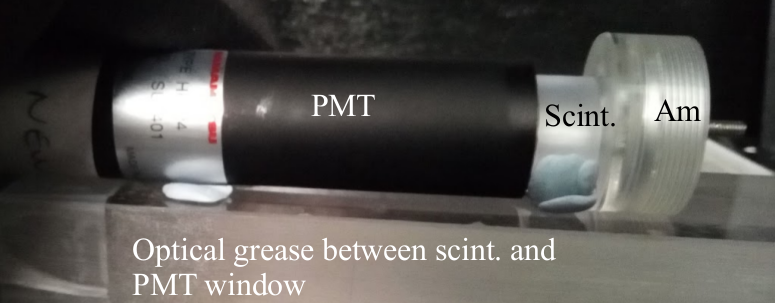}}$%
$\vcenter{\includegraphics[scale=0.35,type=pdf,ext=.pdf,read=.pdf]{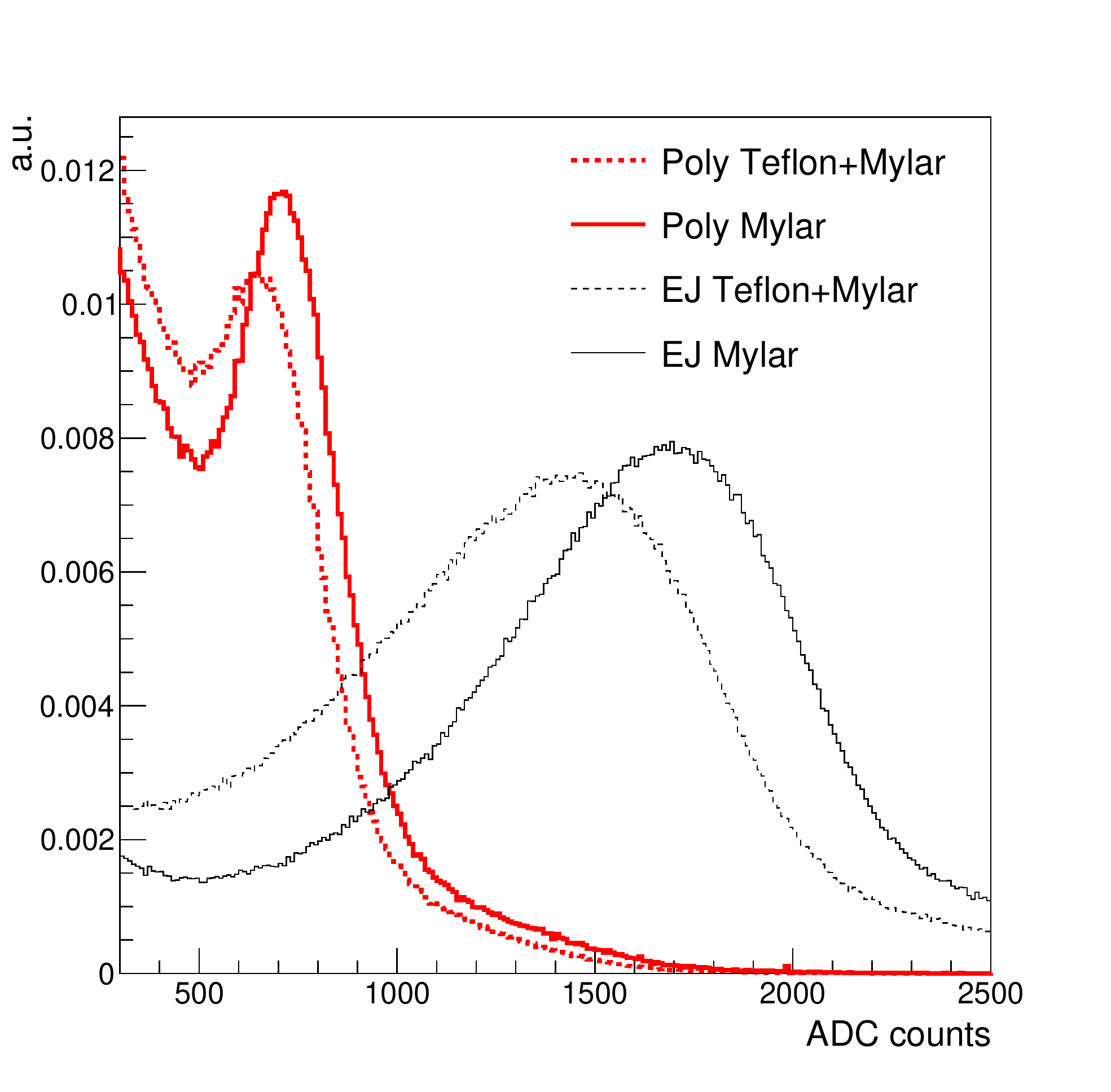}}$
\end{minipage}
\caption{Comparison of light yields with sources. Left: The setup with the Am source, the scintillator cylinder and the photomultiplier. Right: pulse spectra for the Eljen scintillator (thin black) and the polysiloxane one (thick red). Dashed (solid) histograms are obtained with wrapping 1 (2), respectively, see text.}
\label{fig:sources}
\end{figure}

The light yield of our samples of polysiloxane-based scintillators has
been compared to that of a classic plastic scintillator by exposing them to an Am source\footnote{Am decays
  mainly by $\alpha$ emission - 5.486~MeV - and a $\gamma$ - 54~
  keV.}.  The source (Fig.~\ref{fig:sources} left) was put in contact
with cylindrical samples of 1 cm thickness and 2.35~cm diameter made
with polysiloxane or plastic scintillator (surface polished
EJ-200). The other side of the cylinder, i.e. the flatter surface corresponding to the base of the container were the
  polysiloxane cylinder was poured, was put in contact with a
Hamamatsu 2.35~cm diameter photomultiplier (mod. H6524) with a thin
optical grease layer in between to improve the light transmission. The
samples were wrapped to maximize the light output in two ways: 1) with
a reflective thin Mylar foil on the source side and Teflon tape for
the cylinder side or 2) using the same Mylar layer for all
surfaces. The PMT signals were amplified and shaped before being
acquired by a multi-channel analyzer.
%(CAEN {\bf xx}). 
The measured spectra are shown in Fig.~\ref{fig:sources} (right) for
the polysiloxane (thick red) and the Eljen plastic scintillator (thin black) with the two
scintillator wrapping options (1) - dashed - and (2) - dotted. Mylar (2)
gives a better light yield in both cases with a relative increase of
about 20\% with respect to the other wrapping option. The ratio of the
light yields of Eljen and polysiloxane, estimated from the ratio of
the peaks positions in ADC counts (Fig.~\ref{fig:sources} right), is
between 2.2 and 2.4 thus roughly confirming results from the
literature.

\section{Optical simulation}
\label{optisim}
The polysiloxane reticulates around the fibers leaving no air gap between them.  This causes a significant difference in the collection and transmission of photons.
We consider here the setup of Sec.~\ref{setup} in which the polysiloxane is coupled to Kuraray~\cite{ref23} Y11 WLS fibers while the standard option employs the EJ-204~\cite{ref22}
plastic scintillator and Saint-Gobain~\cite{ref24} BCF92 WLS fiber. The conclusions are anyway not critically dependent on the choice of the WLS fiber or the scintillator but rather driven by the presence or absence of the air gap.
The absence of the air layer allows a better collection of photons since the critical angle for a total reflection of photons exiting the plastic scintillator ($n_1=1.58$ for EJ-204) towards the air gap ($n_2=1$) is about 39$^\circ$
while it is about 70$^\circ$ degrees when passing directly from the Polysiloxane ($n_1=1.51$) to the outer
cladding of the Y11 fibers ($n_2=1.42$). In terms of solid angles
($\Omega=2\pi(1-\cos\theta)$) this corresponds to a gain of about a factor three. On the other hand the fiber itself loses trapping efficiency since all the rays reaching the outer
cladding layer pass to the scintillator that has a higher refraction index. 

\begin{figure}[t]
\centering
\includegraphics[scale=0.7,type=pdf,ext=.pdf,read=.pdf]{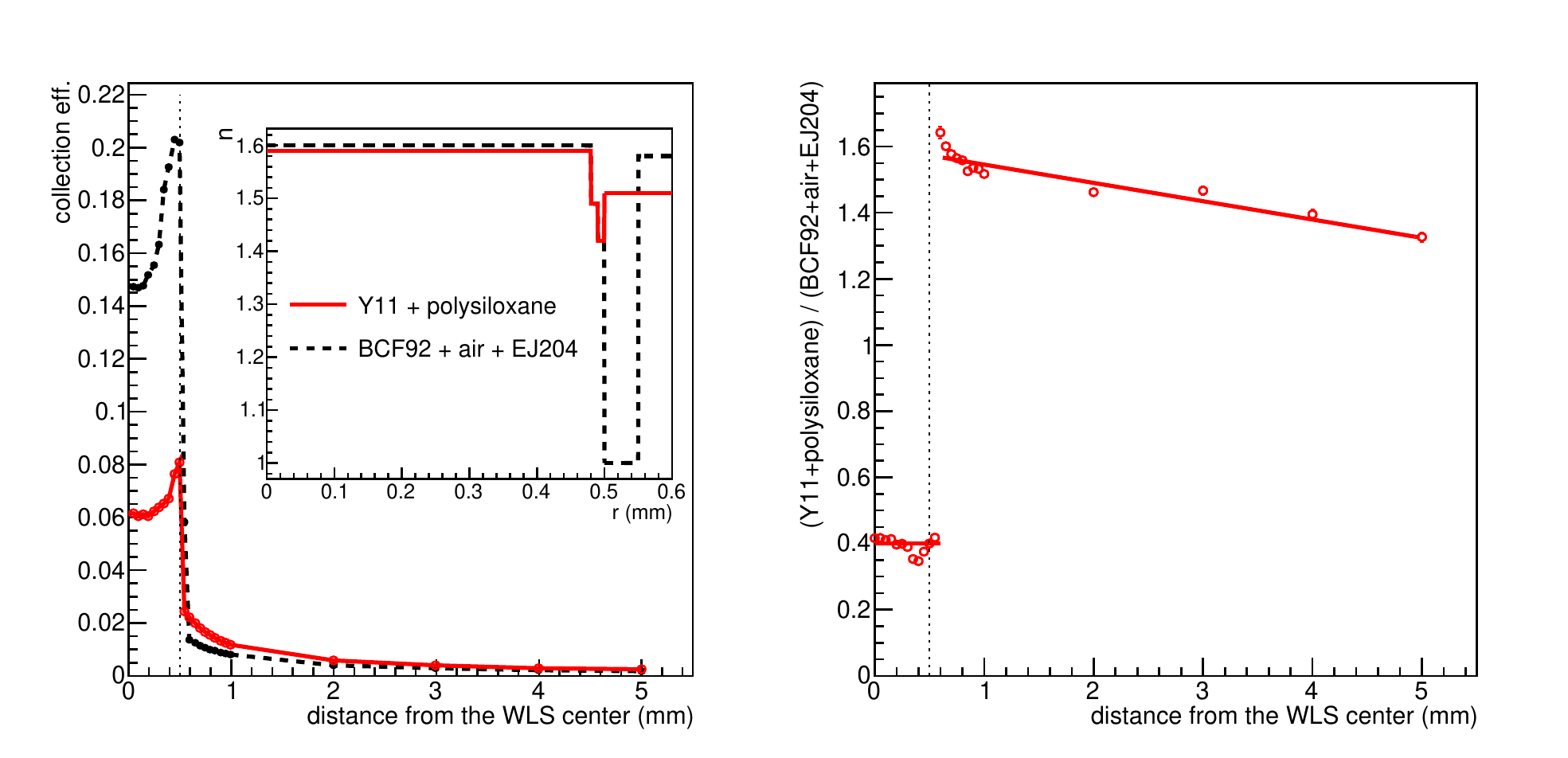}
\caption{GEANT4 optical simulation results.
Left: Collection efficiency as a function of the distance from the WLS center for the Y11+polysiloxane (solid red) and the BCF92+air+EJ204 (dashed black) setups. Right: ratio of the collection efficiencies for the two setups. The dotted vertical line represents the boundary of the WLS fiber. The inset in the left plot shows the refraction index from the fiber center for the polysiloxane (solid red) and the plastic scintillator (dashed black) configurations .The refraction index
  of the outer part of Y11 WLS is, from the core to the outer
  cladding, 1.59, 1.49, 1.42 (1.05, 1.19, 1.43 g/cm$^3$) while that of
  polysiloxane is 1.51 (1.01 g/cm$^3$). The refraction index of EJ-204
  is 1.58 (1.023 g/cm$^3$) while that of BCF-92 is 1.60, 1.49, 1.42
  (1.05 g/cm$^3$ density for the core).  The vertical line represents the boundary of the WLS fiber.}
\label{fig:optisim}
\end{figure}

The interplay of the two effects was estimated using a GEANT4~\cite{geant1,geant2,geant3}
simulation of the multi-clad Y11 fibers with an isotropic source of
optical photons. The source was moved radially from the
center of the fiber and the number of collected photons was
measured in a configuration with a 0.05~mm air gap or no gap. Figure
~\ref{fig:optisim} (left) shows the collection efficiency as a
function of the distance of the source from the fiber in the presence of an air gap
(dashed black) and without an air gap (solid red). The ratio of the two curves (Y11+Polysiloxane / BCF92 + air gap + EJ204)
is shown in the plot on the right. 
It is clear that photons generated
inside the fiber are more likely to return to the scintillator (-60\%)
in the absence of an air gap and that, on the contrary, when emitted in the
scintillator they have a larger chance of making their way up to the
WLS fiber end (about +40\% when averaging over a distance of 5~mm i.e. half the fiber spacing). An overall enhancement of $\sim$~40\% in the light yield is expected in the polysiloxane configuration after normalizing for other effects (i.e. the intrinsic scintillation light yields). This result is compatible with considering a rough factor three in collection according to the considerations described above and the 60\%
reduction in trapping efficiency of the WLS fiber given by the simulation.

\section{Layout and construction of the calorimeter prototypes}
\label{setup}
Before building the 13~$X_0$ thick calorimeter used for this work,
several smaller prototypes were tested in 2016-17 at the
CERN-PS, on the scale of a single UCM (3$\times$3~cm$^2$ with a
4.3~$X_0$ depth).  
The results of these pilot tests
allowed tuning the scintillator thickness and the production
techniques. This prototype will be henceforth denoted as ``POLY''
(Sec.~\ref{POLY}) while we will denote with ``PLAS'' the reference
calorimeter composed of plastic scintillators (Sec.~\ref{setupref}).

\subsection{Preparation of the polysiloxane prototype}
\label{POLY}

\begin{figure}
	\centering
	\includegraphics[scale=0.13,type=png,ext=.png,read=.png]{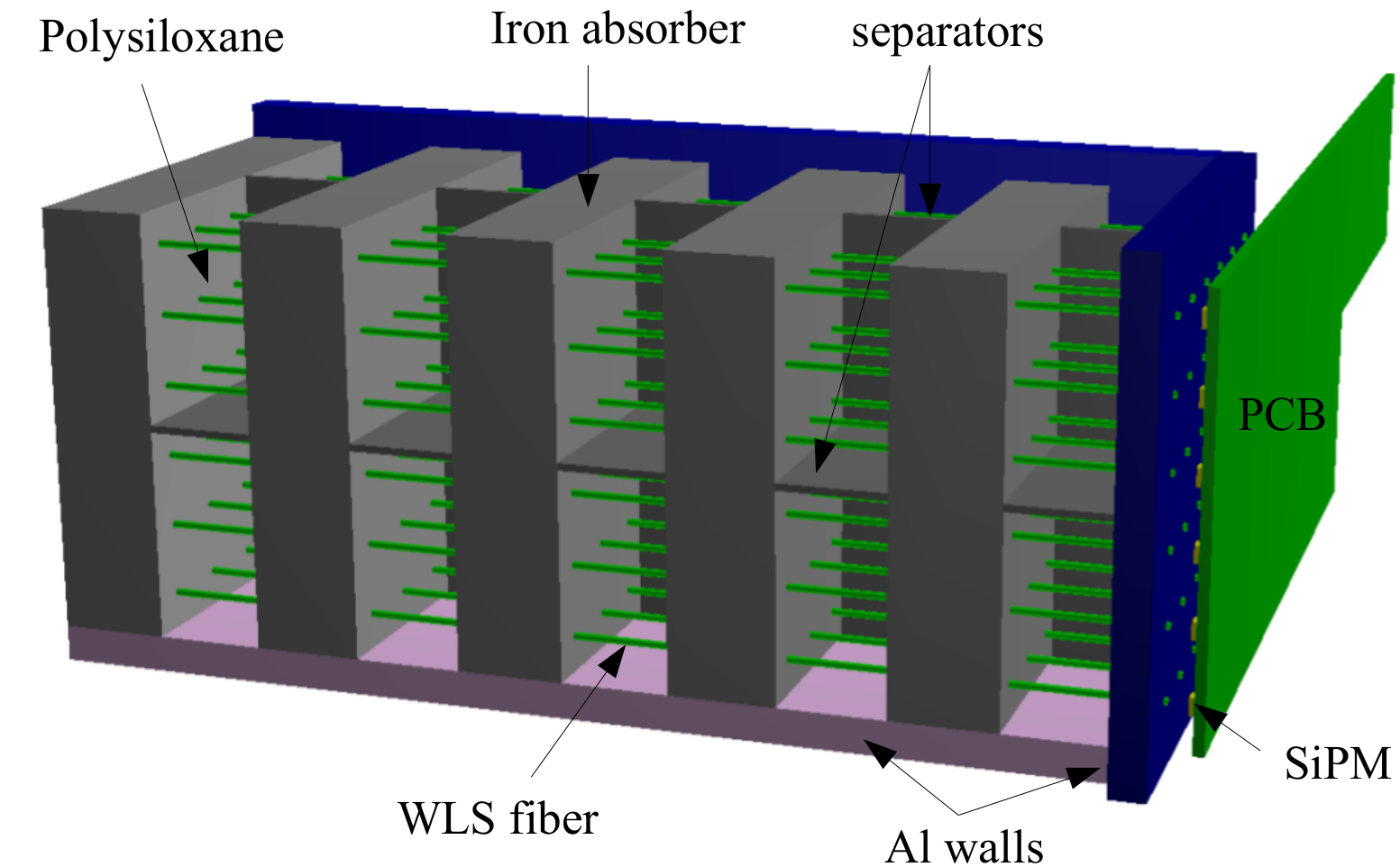}%
	{~~~~~~}%
	\includegraphics[scale=0.14,type=png,ext=.png,read=.png]{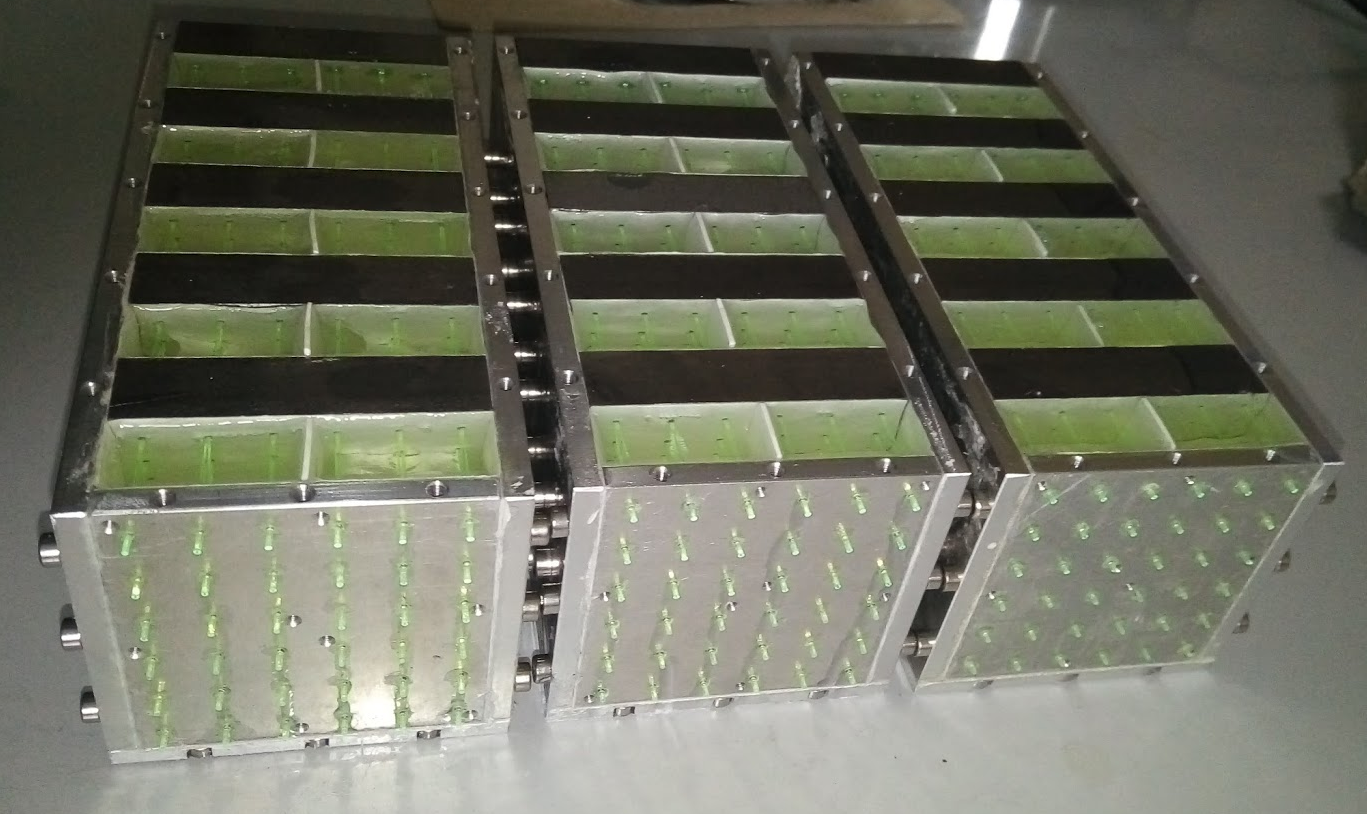}
	\caption{Polysiloxane prototype. Left: annotated sketch of a single module (4~UCMs). Right: a picture of the three calorimeter modules (12 UCMs).}
	\label{fig:12ucm}
\end{figure}

Each of the three modules composing the POLY calorimeter
(Fig.~\ref{fig:12ucm} left) consists of 4~UCMs: 2$\times$2 in the
plane perpendicular to the beam, with a transverse size of 3$\times$3 cm$^2$ each.  The calorimeter is then composed by
12 UCM (2$\times$2$\times$3) with iron and scintillator layers 1.5~cm
thick. The light produced in the scintillator is read out by 9 WLS
fibers/UCM (diameter: 1~mm, length: 15~cm) and each UCM is composed by
five tiles: it thus samples 4.3~$X_0$ along the development of the
shower and 1.7~Moli\`ere radii in the transverse plane.  Transversally
the iron slabs are 6$\times$6~cm$^2$ wide.  The slabs were drilled
with a CNC machine: the distance between holes is 1~cm and the
diameter of the holes is 1.2~$\pm$~0.2~mm. The absorbers were hosted
inside a liquid-tight 5~mm thick aluminum structure to which they were
fixed with bolts.
A Tyvek\textsuperscript{\textregistered} foil
was glued to the absorber sides facing the scintillator and the inner
sides of the Al container, using a cyanoacrylate resin.
Holes in Tyvek were made with a hot needle by using the
pattern of holes in the absorber as a guide, after they had been fixed
on the absorbers' surfaces.

The optical separation between the 4~UCM in each module was achieved
by inserting a cross-shaped white polystyrene septum with 1.5~cm
depth and 1~mm thickness. No Tyvek or TiO painting was hence applied
to the separator. A pair of 1~mm diameter holes was cast in the
separator in the middle of each UCM to allow the flow of liquid
polysiloxane from the top to the bottom pair during the pouring phase.

\begin{figure}
  \centering
  \includegraphics[scale=0.121,type=jpg,ext=.jpeg,read=.jpeg]{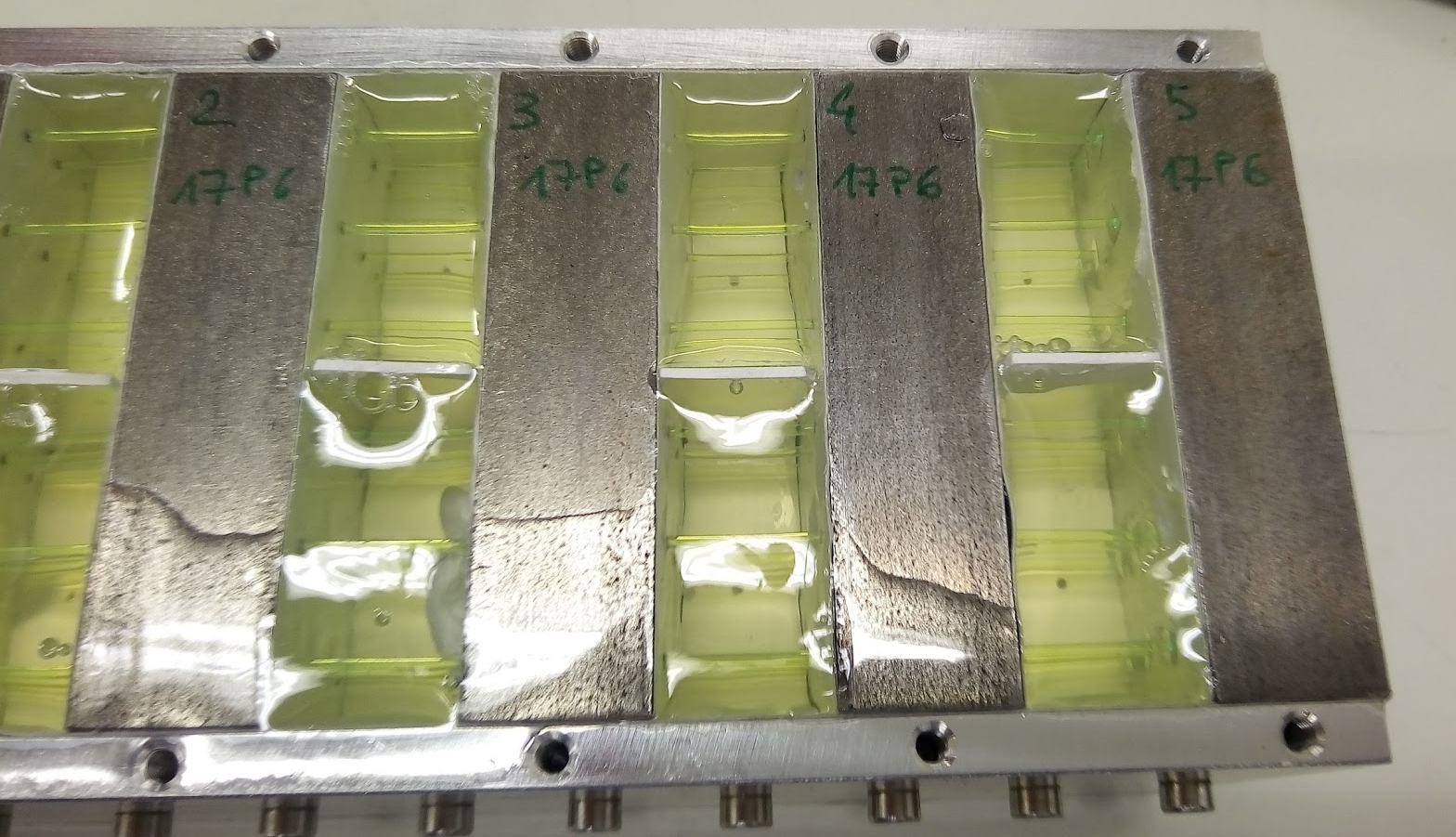}%
                  {~~~~}%
  \includegraphics[scale=0.114,type=jpg,ext=.jpeg,read=.jpeg]{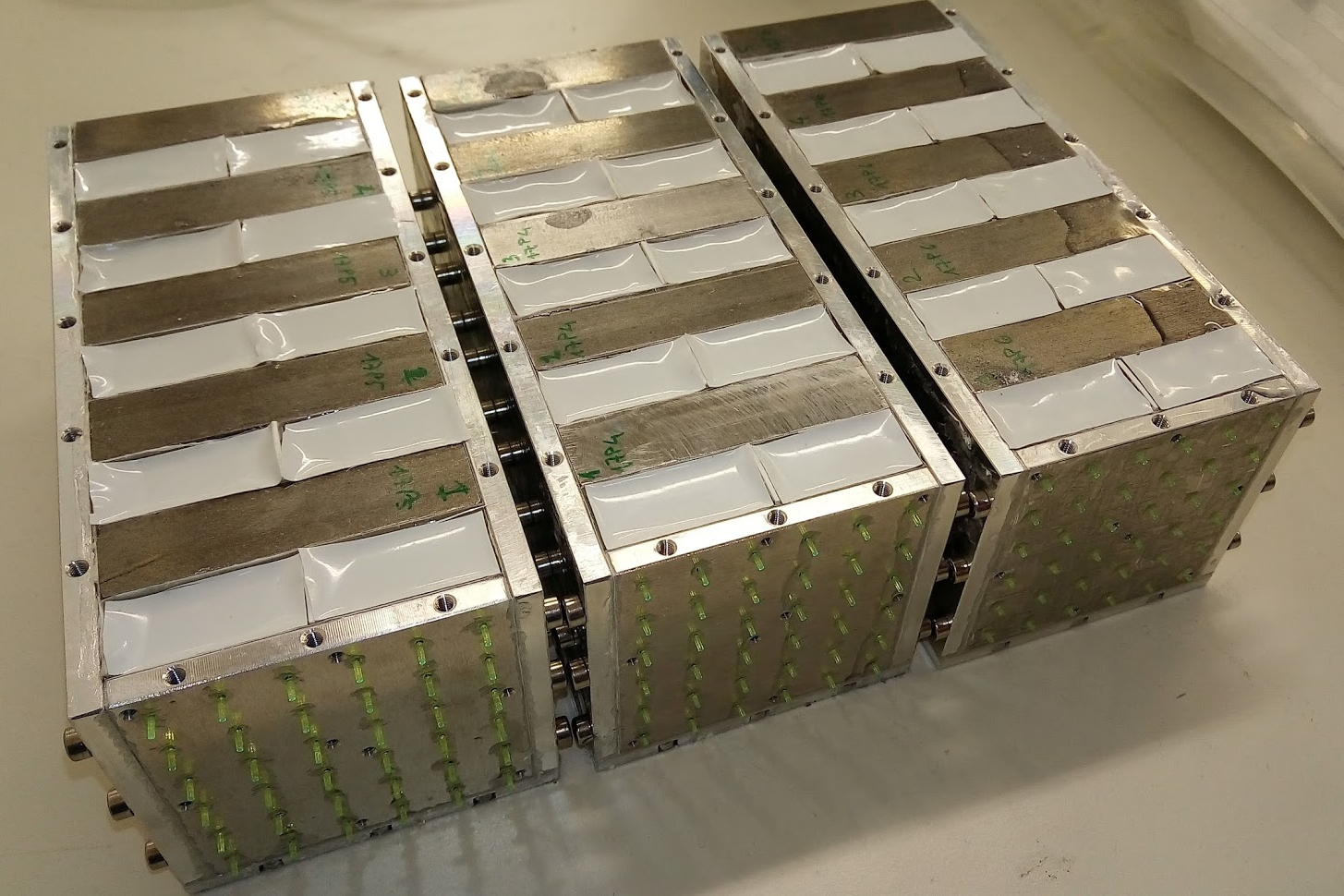}
  \caption{Polysiloxane prototype. Left: close up of
    scintillator compartments. Right: the full prototype with a
    white diffuser on the surface face.}
  \label{fig:WGpics}
\end{figure}

The fibers used are Kuraray Y11
multi-clad with a 1~mm diameter. The
WLS fibers were polished with fine
sand paper foils on a rotating polisher and an Al mirror was placed on the side opposite to the SiPM and
attached to the WLS polystyrene using a cyanoacrylate
resin (visible in Fig.~\ref{fig:WGpics} left).

Since the WLS fibers could move longitudinally before the reticulation
of polysiloxane, particular care had to be put into equalizing the WLS
longitudinal position at the SiPM side to keep them as planar
as possible. 

The polysiloxane mixture we used contains vinyl terminated
polymethyl-phenyl siloxane as precursor, which is cross-linked with a
Si-H containing resin and a suitable amount of Pt-based
catalyst. Additives to achieve good light output, meanwhile preserving
transparency, are 2.5-diphenyl oxazole (PPO, Sigma Adrich) and Lumoen
Violet (LV, Basf), added in 1\% and 0.02\% respectively. This
composition allows to obtain the best performance as for light output,
as shown in Refs.~\cite{degerlier,Quaranta2010b}.
The viscous resin was degassed in vacuum (0.1~mbar) and then poured
with a syringe. The prototype was put into an oven at a temperature of
60$^\circ$C for 24 hours.

In spite of outgassing, the occurrence of bubbles could not be fully
avoided, since the pouring procedure through the syringe causes air trapping. During reticulation, bubbles migrate towards the
top of the polysiloxane layer and remain trapped therein, as visible in Fig.~\ref{fig:WGpics} left. Moreover, the formation of discontinuity planes inside the scintillator where light gets totally reflected was also observed, probably resulting from mechanical stresses in the cooling phase. An example is visible again in Fig.~\ref{fig:WGpics},
left, in the second scintillator from left, bottom row.  Since the top
surface of the polysiloxane is not perfectly planar after reticulation
(Fig.~\ref{fig:WGpics} left), it was covered with a white diffusive
tape which followed the curvature of the surface and minimizes light
losses (Fig.~\ref{fig:WGpics} right).

The array of SiPMs reading the UCM is hosted on a PCB (Printed Circuit
Board) holder that integrates both the passive components and the
signal routing toward the front-end electronics.
This scheme combines the compactness of SiPM-based
calorimeters~\cite{ref19} with the flexibility offered by the shashlik
technology in choosing the longitudinal sampling (length of the fiber
crossing the scintillator/absorber tiles) and transverse granularity
(tile size, number of fibers per unit surface and number of summed
SiPM channels).

\begin{figure}
\centering
\includegraphics[scale=0.11,type=png,ext=.png,read=.png]{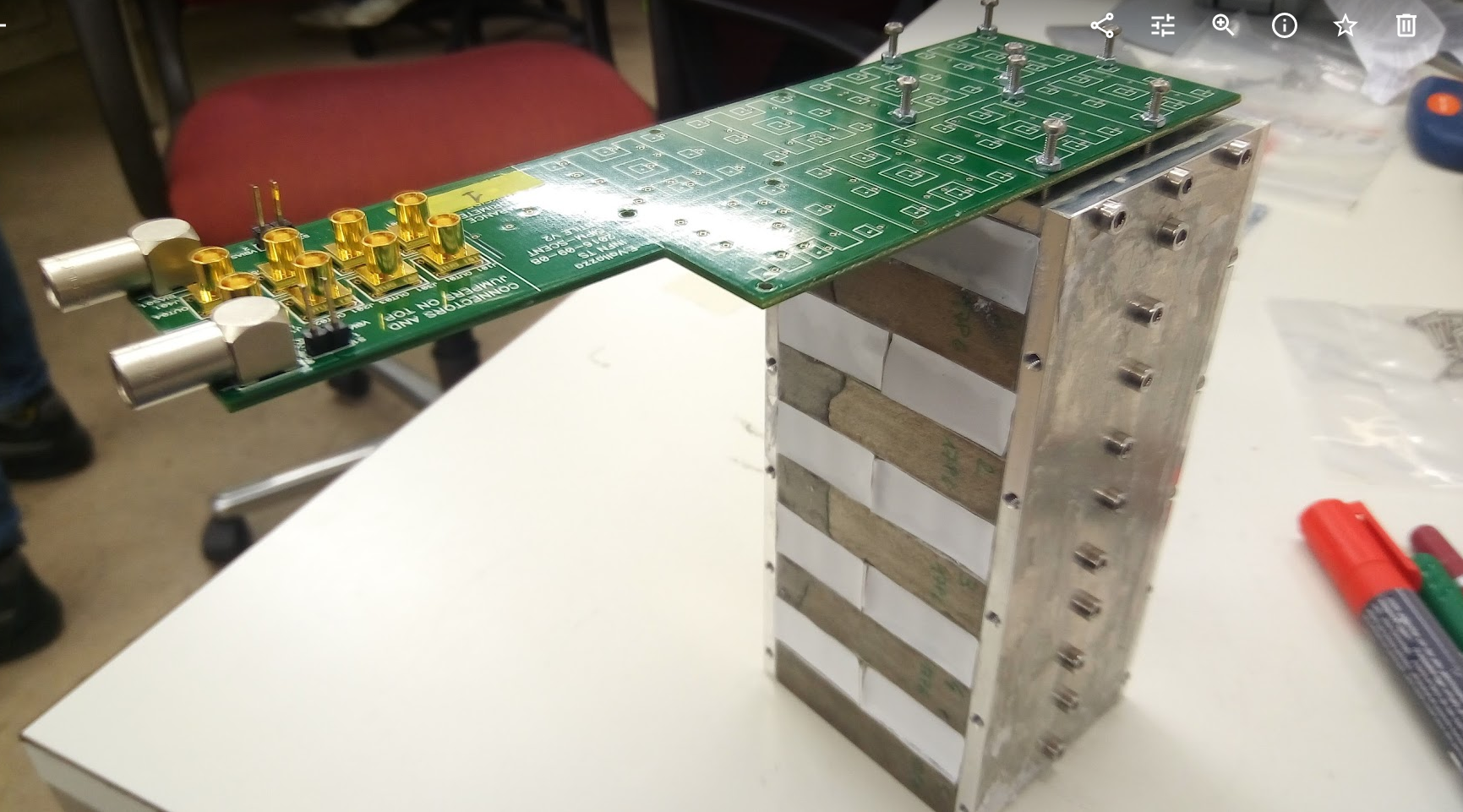}%
                {~~}%
\includegraphics[scale=0.212,type=png,ext=.png,read=.png]{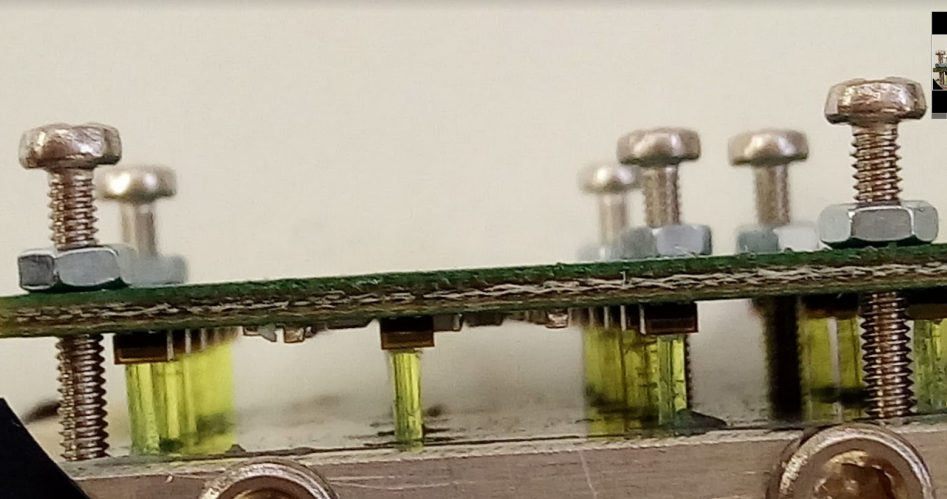}
\caption{WLS-SiPM coupling. Left: a single module with the PCB
  installed in the back plane. Right: a close-up lateral view of the
  PCB with fixing bolts. The Y11 WLS fibers touching the SiPM are
  visible and could be individually inspected.}
\label{fig:coupling}
\end{figure}

The coupling of the WLS fibers to the SiPM matrix embedded in the PCB
is shown in Fig.~\ref{fig:coupling}. In a first
version of the calorimeter the WLS fibers were cut 7~mm above the downstream aluminum face. In this way it was possible to check the good contact between the WLS and the SiPM
by visual inspection. The alignment of the fibers and SiPM was
ensured by 7 screws connecting the PCB and the aluminum layer.
Due to mechanical stresses though we noted a slight bending of the WLS fibers. The bending introduces a visible lateral displacement between the
SiPM and the fibers. In a later version of the prototype the most upstream module was
improved and the fibers brought to the same level of the aluminum
(Sec.~\ref{ly}).

The light transmitted by the fibers is read by 1 mm$^2$ SiPMs with
20~$\mu$m cell size. The sensors are developed by FBK~\cite{ref25} and
are based on the $n$-on-$p$ RGB-HD technology. Each SiPM hosts 2500
cells in a 1~mm$^2$ square with a fill factor
of~66\%~\cite{ref26}. Each module, corresponding to 4~UCMs, requires
9$\times$4~=~36 SiPMs.  The calorimeter (3~modules) hence hosts
108~SiPMs. These SiPMs were produced by FBK from a single wafer and
encapsulated in a chip-scale epoxy package (SMD package) by Advansid
s.r.l.~\cite{ref27}. The epoxy layer between the silicon layer and the
WLS has a thickness of 200~$\mu$m.
The $V$-$I$ response was characterized at the production site. Since
all SiPMs of the calorimeter were produced starting from the same
silicon wafer and in a single lot, the breakdown voltage is very
uniform among the sensors: 28.2~$\pm$~0.1~V.  The SiPMs are mounted as
standard SMD components on a custom 6-layer PCB hosting all the
sensors belonging to the same module.

The SiPMs belonging to the same UCM are connected in parallel and read
out without amplification through a 47~pF decoupling capacitor. The
PCB is equipped with a flap hosting 8 MCX connectors to read the
signal of the UCMs. In the PCBs used for the calorimeter the bias is
the same for all SiPMs and it is distributed by a coaxial cable. Each
PCB hosts 72 SiPM so only half were biased and used for this
measurement.

\subsection{The reference calorimeter with standard scintillators}
\label{setupref}
\begin{figure}
  \centering
  \includegraphics[scale=0.25,type=png,ext=.png,read=.png,angle=90]{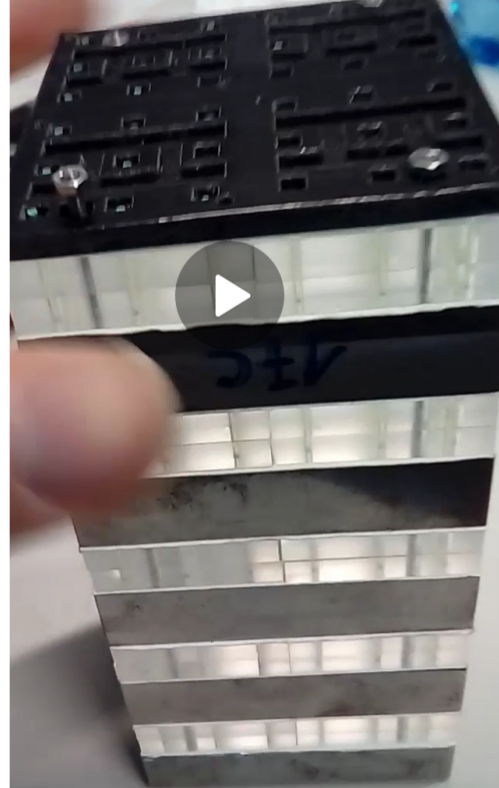}%
  {~~~~}%
  \includegraphics[scale=0.14,type=png,ext=.png,read=.png,angle=0]{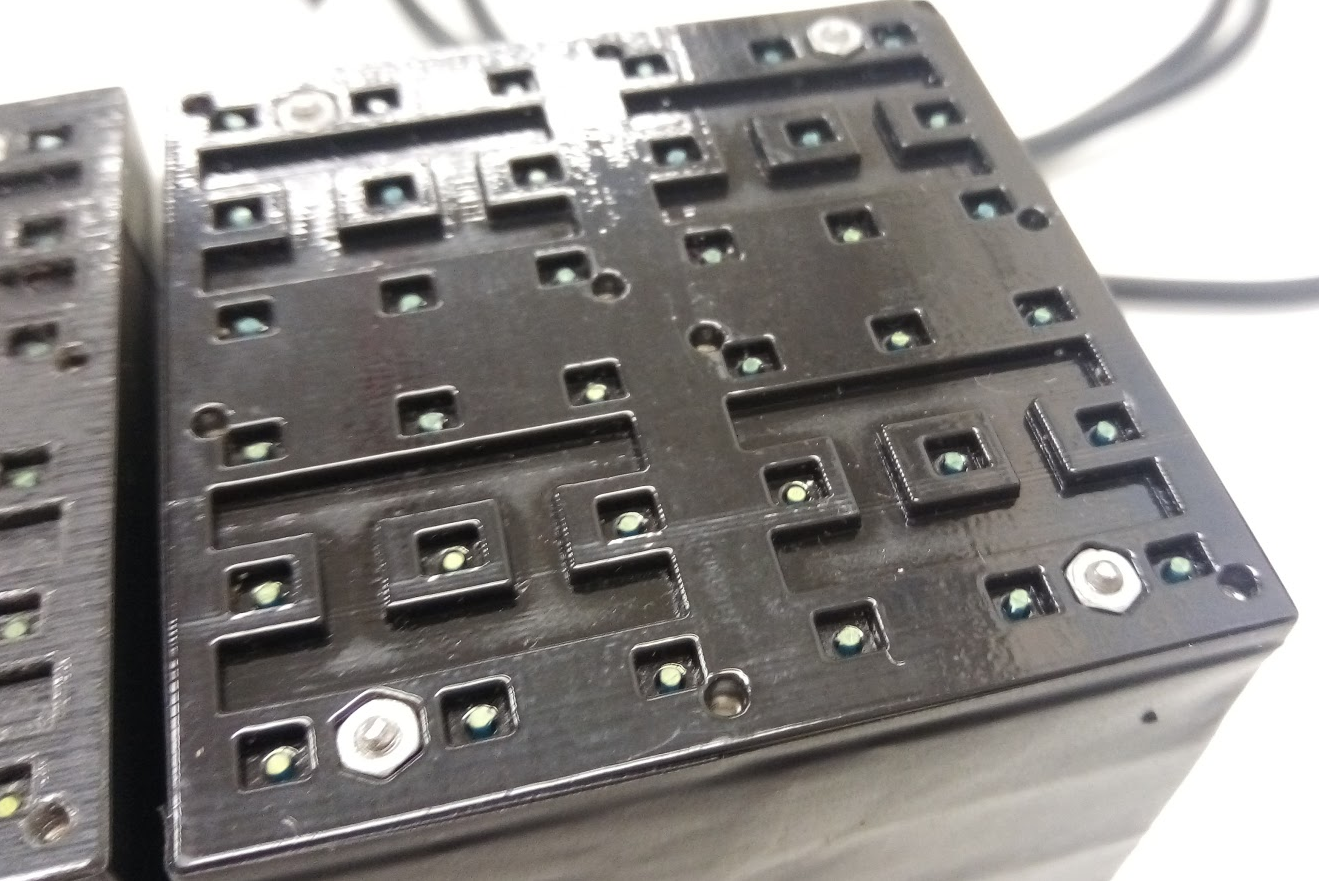}
  \caption{PLAS reference calorimeter. Left: the
    absorber--scintillator stack before wrapping with black
    tape. Right: a detail on the 3D-printed mask used to
    interface the WLS fibers and the SiPM mounted on the PCB.}
  \label{fig:FF}
\end{figure}

The shashlik calorimeter with plastic scintillators used for
benchmarking (PLAS) is shown in Fig.~\ref{fig:FF}.  The scintillator
tiles (3$\times$3~cm$^2$, thickness 0.5~cm) were machined and
polished using EJ-204 plastic scintillator sheets. Each
layer is composed of a pair of such tiles for a total thickness of
1~cm.  Holes were drilled (up to 4 tiles per stack) with a 1 cm pace
and a $1.2\pm 0.1$~mm diameter using a CNC machine with controlled
rotation speed to prevent distortions caused by heating. Tyvek foils
were used in between the scintillator and the absorber layers while
all lateral sides were covered with a Mylar reflective foil before
wrapping the stack with black tape.

The 1 mm diameter WLS fibers (Saint Gobain BCF92
multi-clad) cross the modules through the holes up to the last
scintillator plane and are connected to a 3D printed plastic mask
located downstream of the module, as shown in Fig. \ref{fig:FF},
right. The plastic mask is grooved in order to fix the fibers in the
back of the module and couple them with the PCB hosting the same
SiPMs used for the POLY calorimeter. Four threaded bolts (2 mm diameter) cross the module and are
fixed to the plastic mask by nuts positioned into the mask itself.
BCF-92 fibers offer a fast response (2.7~ns) compared with Y11 (10~ns).
The SiPMs are aligned to the fibers in the transverse plane with a
precision of 0.1~mm via the mechanical coupling of the PCB with the
plastic mask.

This calorimeter too is composed by 12 UCM (2$\times$2$\times$3) for a total thickness of about 13 $X_{0}$.
A summary of the parameters of POLY and PLAS is given in
Tab.~\ref{tab:sum}.
\begin{table}
  \centering
  \begin{tabular}{|c|c|c|c|c|}
\hline
 prototype & scint. thick. (mm)& abs. thick (mm)& scint. & WLS \\
\hline
POLY & 15 & 15 & polysiloxane & Y11 \\
\hline
PLAS & 10 & 15 & EJ204 & BCF92 \\
\hline
  \end{tabular}
  \caption{Summary of the parameters of POLY and PLAS.}
\label{tab:sum}
\end{table}

\begin{figure}
	\centering
	\includegraphics[scale=0.68,type=pdf,ext=.pdf,read=.pdf]{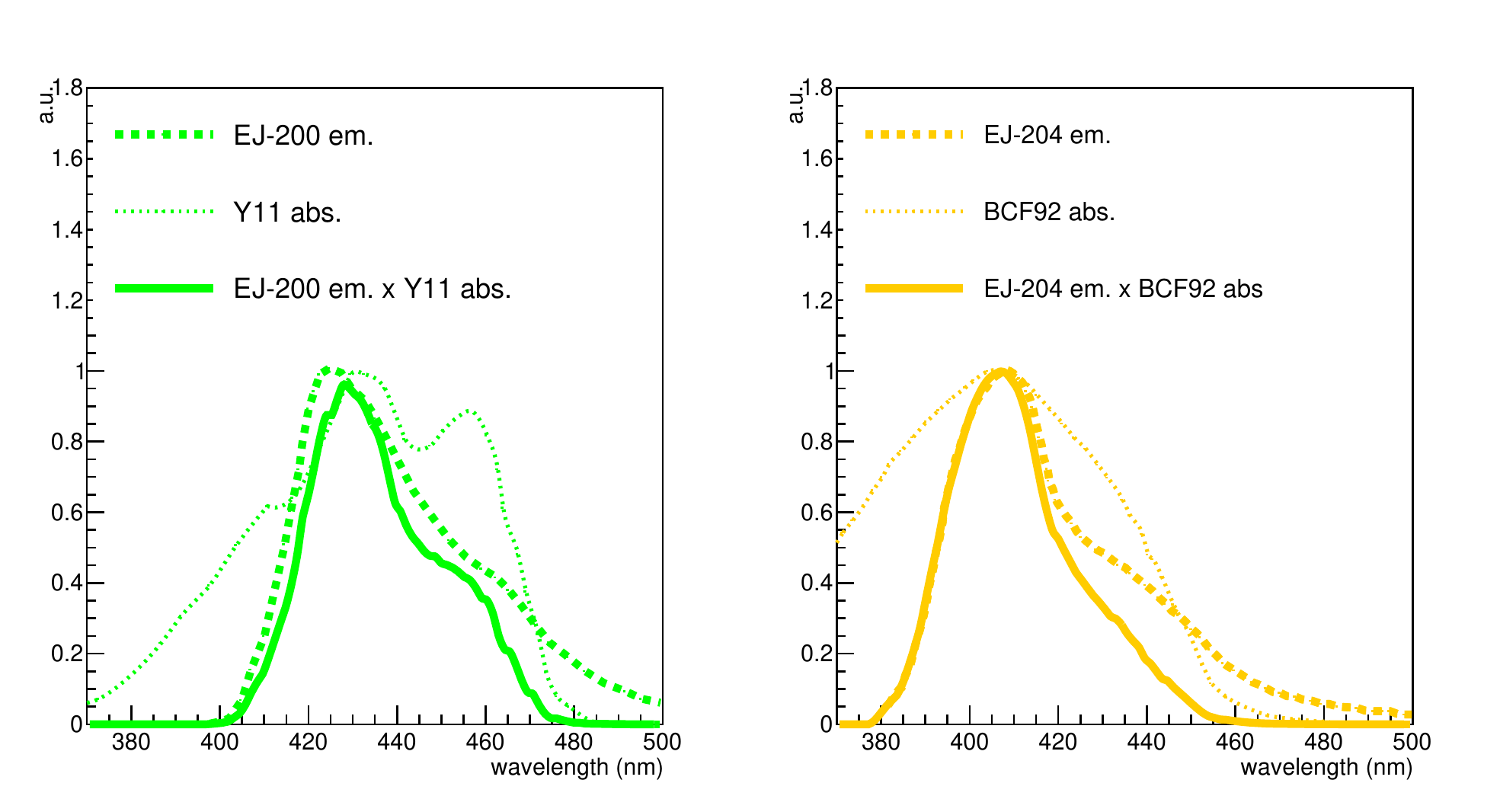}
	\caption{Right: spectral matching between BCF92 fibers and EJ-204 scintillator, equivalent to the PLAS setup. Left: spectral matching between Y11 fibers and EJ-200, corresponding to the POLY setup (EJ-200 emission is very similar to the used polysiloxane-based scintillator).}
	\label{fig:spectralmatch}
\end{figure}

We have evaluated the integral of the product of the
scintillation light spectrum and the spectral absorption profile of
the fibers for the two setups,
and they do not differ by more than 5\% (see Fig.~\ref{fig:spectralmatch}).

\subsection{Test setup in the PS-T9 beamline}
\label{t9}
The calorimeters were exposed to electrons, muons and pions at the CERN
PS East Area facility for two weeks in October 2017 and May 2018. The
momentum of the particles was varied between 0.5 and 7 GeV. The
detector was positioned inside an aluminum box to ensure light
tightness and mounted on a platform in the T9 experimental area in
front of two silicon strip detectors.  The silicon
detectors~\cite{ref29, ref31} provide track reconstruction with a
spatial resolution of 30 $\mu$m.  A pair of threshold Cherenkov
counters filled with CO$_2$ are located upstream of the silicon
detectors.  The maximum operation pressure of the counters is 2.5 bar:
they were thus used to separate electrons from heavier particles
($\mu$ or $\pi$) below 3 GeV and the muon/pion separation was obtained using a muon catcher located downstream of the calorimeter. Between 3~and~5~GeV the two counters
were operated at different pressures to identify electrons, muons and
pions. A 10$\times$10~cm$^2$ plastic scintillator located between
the silicon and Cherenkov detectors is employed as trigger for the
DAQ. Particles in the beamline are produced from the interaction of
the primary 24 GeV/c protons of the CERN-PS accelerator with a fixed
target. During the test, we employed the T9 ``electron enriched''
target. It consists of an aluminum tungsten target (3
$\times$5$\times$100~cm$^2$) followed by a tungsten cylinder
(diameter: 10~cm, length: 3~cm). We set the collimators in order to
provide a momentum bite of 1\%. At 3 GeV the beam composition as
measured by the Cherenkov counters is 9\% electrons, 14\% muons and
77\% hadrons. We only selected negative particles in the beamline and
the contamination of protons and kaons is thus negligible.
During the testbeam all the SiPMs were biased with different voltages
ranging from 32 to 38~V. The response of different UCMs was equalized offline 
using minimum ionizing particle signals as a reference. The signals from the
UCM are recorded by a set of 8 channel v1720 CAEN~\cite{ref28}
digitizers (12 bit, 250 MS/s, 2~V range). Additional details on the setup
and the DAQ can be found in~\cite{papernov2016}.

\section{Results from the CERN beam exposure and with cosmic rays}
\label{results}
We present results in terms of
particle identification (Sec.~\ref{pid}), energy resolution, linearity
(Sec.~\ref{eres}) and spatial uniformity (efficiency maps,
Sec.~\ref{effmaps}). We have also measured the light yield in terms of
photo-electrons per minimum ionizing particle (mip) 
with an improved optical matching between the WLS and SiPMs (Sec.~\ref{ly}).

\subsection{Particle identification}
\label{pid}
\begin{figure}
	\centering
	\includegraphics[scale=0.37,type=pdf,ext=.pdf,read=.pdf]{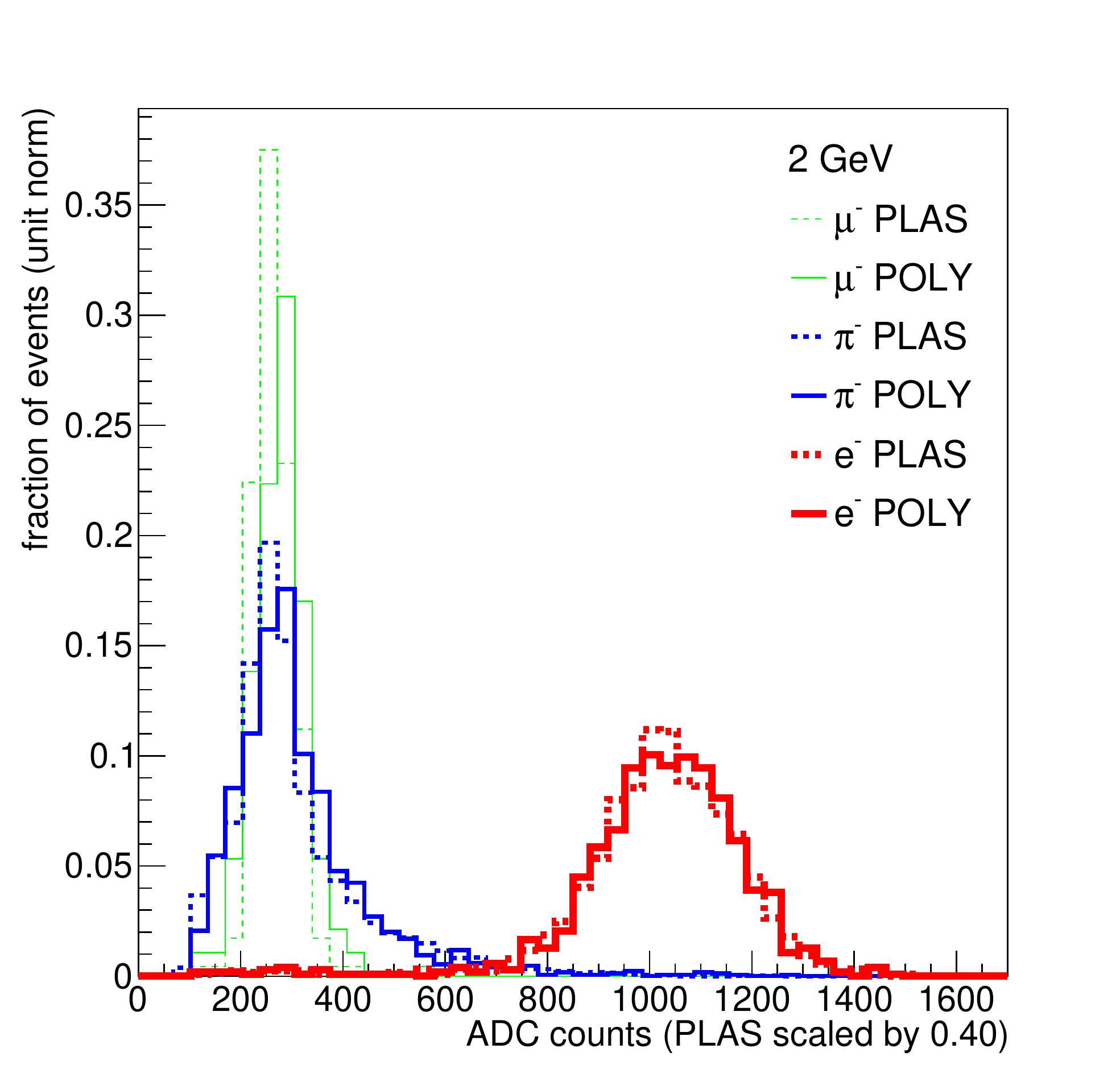}%
	\includegraphics[scale=0.37,type=pdf,ext=.pdf,read=.pdf]{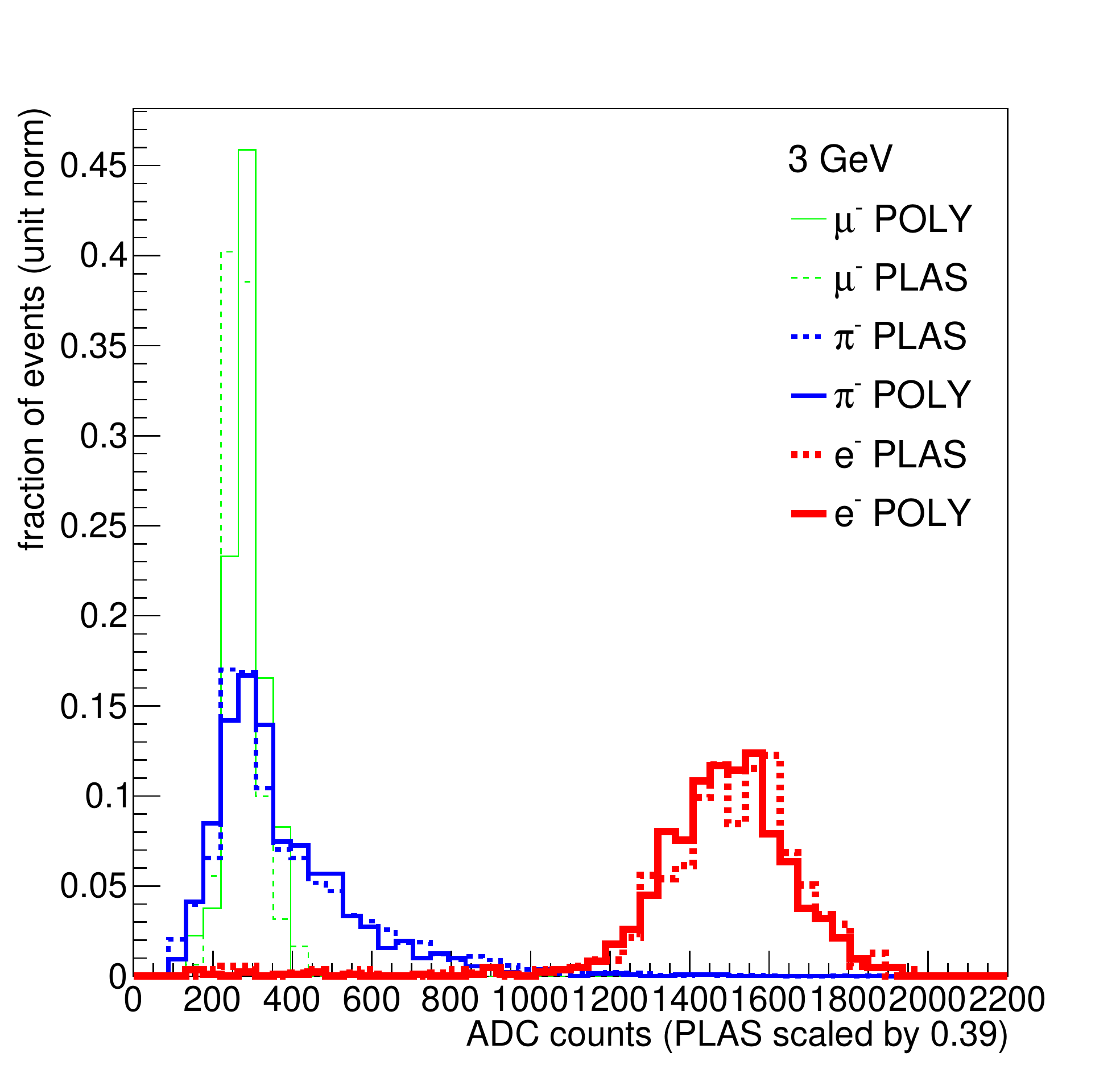}
	\caption{Distributions of the pulse
		heights for POLY (solid) and PLAS (dashed) at a SiPM bias voltage of
		36~V for different particle populations (thicker red line for $e^-$, medium thick blue for $\pi^-$ and thinner
		green for $\mu^-$) as selected using the Cherenkov counters pair. Left and right plots refer to a beam energy of 2 and 3~GeV respectively.}
	\label{fig:WGepi2-3_36bis}
\end{figure}

Figure~\ref{fig:WGepi2-3_36bis} shows the distribution of the pulse
heights for POLY (solid) and PLAS (dashed) at a SiPM voltage bias of
36~V for different particle populations (thicker red line for $e^-$, medium thick blue for $\pi^-$ and thinner
green for $\mu^-$) as selected using the Cherenkov counters and the muon catcher. All distributions are
unit normalized.  For this result and the following, only events with
a single cluster in both silicon chambers were selected.
Left and right distributions refer to a beam energy of 2 and 3~GeV
respectively. Only particles
hitting the calorimeter in a squared fiducial volume of
3$\times$3~cm$^2$ centered in the calorimeter were selected to avoid lateral leakage.

For this comparison the pulse heights of PLAS have been scaled down by
a constant factor of 0.40 and 0.39 for 2 and 3 GeV data respectively.
In spite of a 2.5 smaller signal, the
polysiloxane $e/\pi$ separation performance is very similar to the one
obtained with the standard calorimeter.

The signal pulse height results from the number
of scintillation photons (proportional to the scintillator thickness, $t$,
and the intrinsic light yield, $ly$), the efficiency of the WLS-scintillator
collection ($\epsilon_{coll}$) and the quality of the coupling between the
WLS and the SiPM ($\epsilon_{WS}$).
\begin{equation}
S \propto t \times ly \times \epsilon_{coll} \times \epsilon_{WS}
\end{equation}
In the prototypes under test $t_{POLY} = 1.5~t_{PLAS}$
(15~vs~10~mm) and $\epsilon_{coll, POLY}=1.4~\epsilon_{coll, PLAS}$ (Sec.~\ref{optisim}). The $\alpha$ source measurements provide
${ly}_{PLAS} = 2.4~{ly}_{POLY}$ (Sec.~\ref{sources}). Hence:
\begin{equation}
\frac{\epsilon_{WS,POLY}}{\epsilon_{WS,PLAS}}=\frac{S_{POLY}}{S_{PLAS}}\frac{t_{PLAS} ~ly_{PLAS}~\epsilon_{coll, PLAS}}{t_{POLY}~ly_{POLY}~\epsilon_{coll, POLY}}=\frac{0.4\times 2.4}{1.5 \times 1.4} \simeq 0.5.
\end{equation}
We ascribe the difference mostly to the SiPM-fiber coupling (Fig.5 and 6), which was significantly poorer in POLY than PLAS and has been improved at a later time (see Sec. 7.4).

\subsection{Energy resolution and linearity}
\label{eres}
\begin{figure}
  \centering
  \includegraphics[scale=0.35,type=pdf,ext=.pdf,read=.pdf]{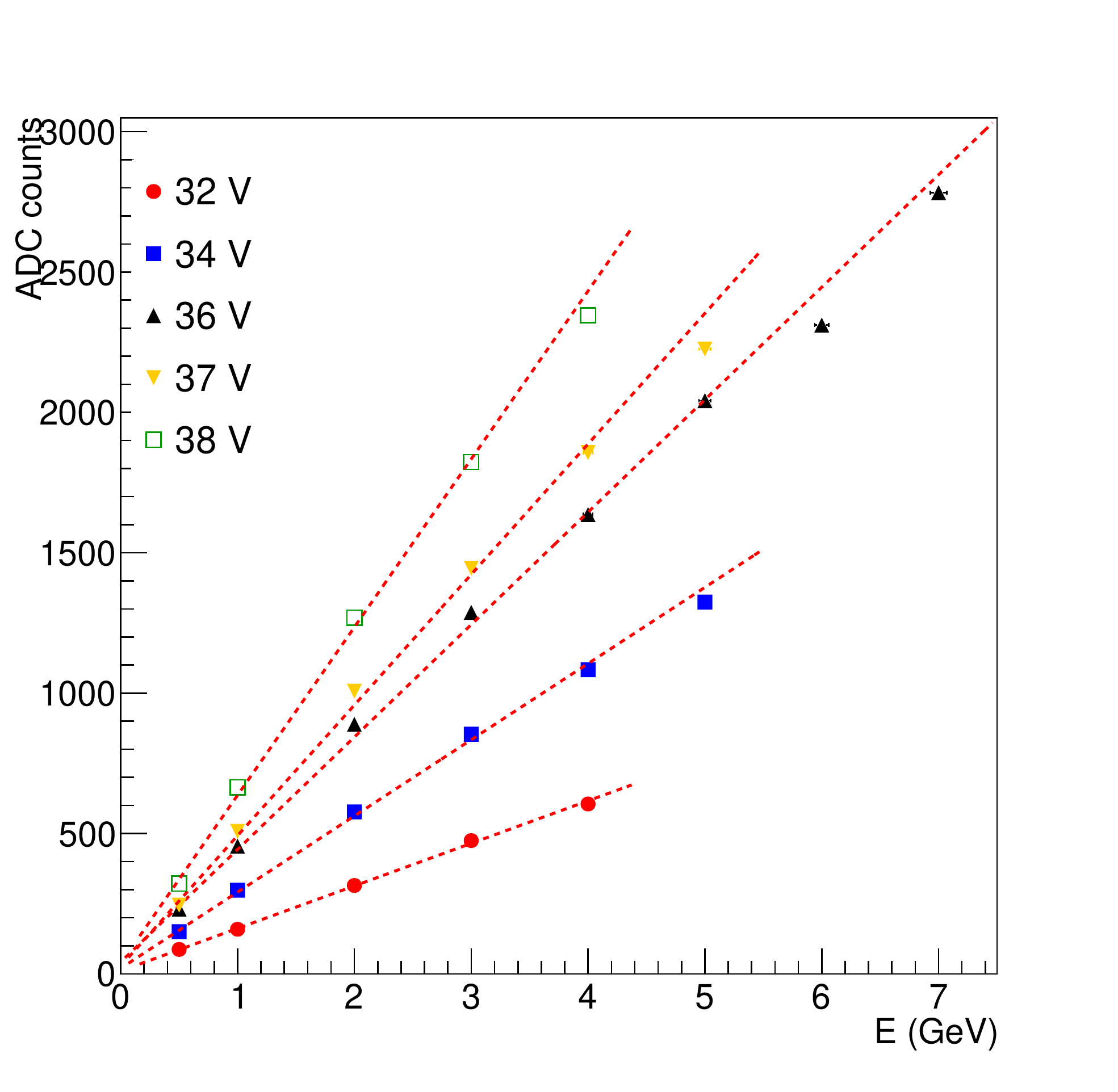}%
  \includegraphics[scale=0.35,type=pdf,ext=.pdf,read=.pdf]{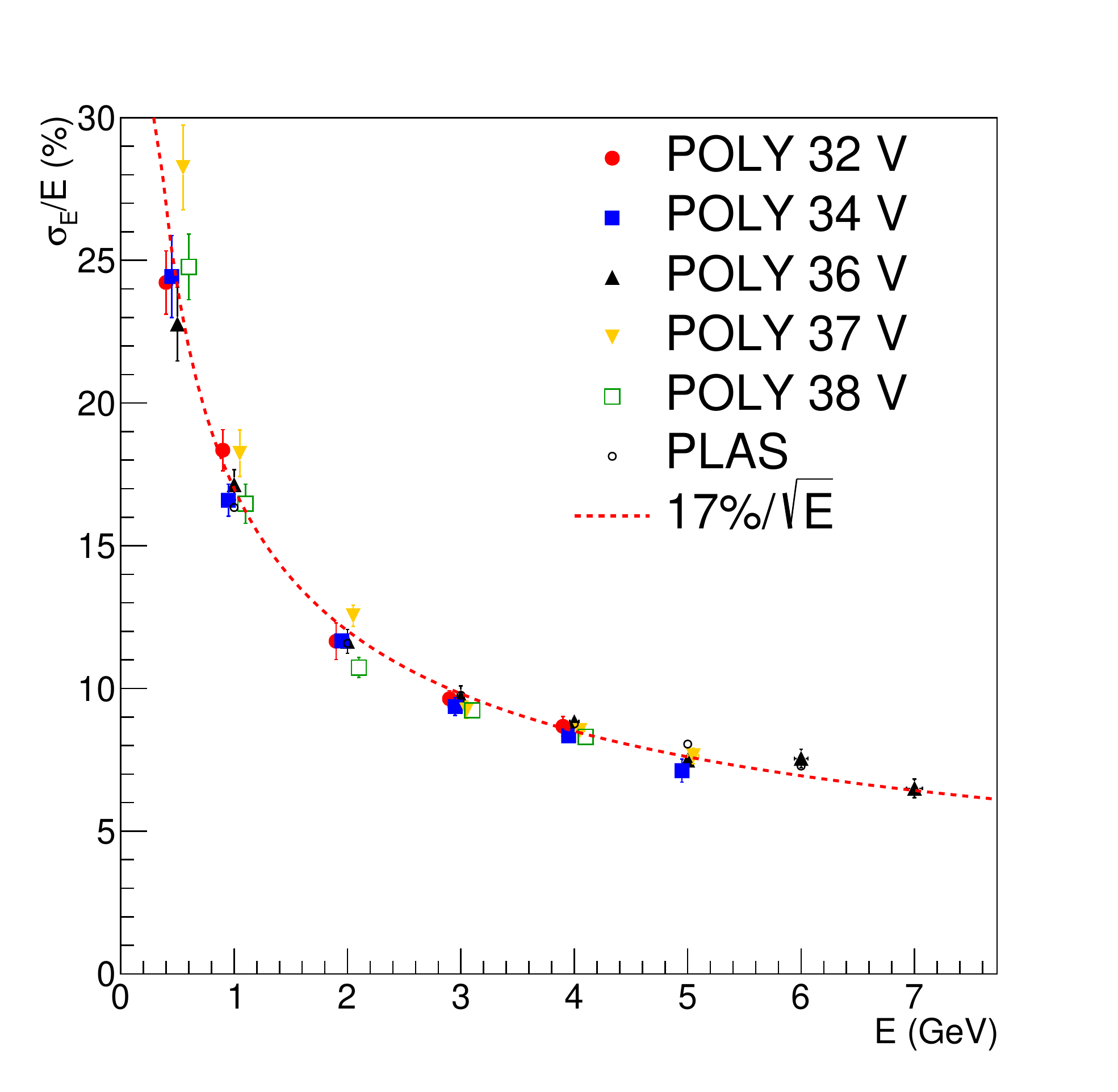}
  \caption{Linearity with electron energy (left) and electron energy resolution (right) for POLY at different values of the SiPM applied voltage.}
  \label{lin0}
\end{figure}

The plot of Fig.~\ref{lin0} (left) shows the dependence of the
position of the electron peak in ADC counts\footnote{An ADC count
  corresponds to a value of 0.488~mV.} (red histograms of Fig.~\ref{fig:WGepi2-3_36bis})
as a function of the beam energy for different bias voltages. Linear
fits are superimposed showing a good linearity with an indication for
a deviation at energies above 4 GeV.

In Fig.~\ref{lin0} (right) the energy resolution, $\sigma/E$, with $\sigma$ being the standard deviation and E the mean value of the Gaussian fits of the pulse height distributions, is
shown as a function of beam energy for different choices of the
bias voltage. The results show no significant
dependence on the bias voltage. Data are well described by a
parametrization of the form $\sigma_E/E = 17\%/\sqrt{E (\rm{GeV})}$ as expected
from the sampling fraction and geometry of the calorimeter. The same
parametrization also describes the energy resolution obtained with
PLAS (Fig.~\ref{lin0} right, hollow round markers).
Hence, the energy resolution of both POLY and PLAS are dominated by the sampling term~\cite{wigmans} more than the collected photon statistics.

\subsection{Efficiency maps}
\label{effmaps}

\begin{figure}
	\centering
        \includegraphics[scale=0.35,type=png,ext=.png,read=.png]{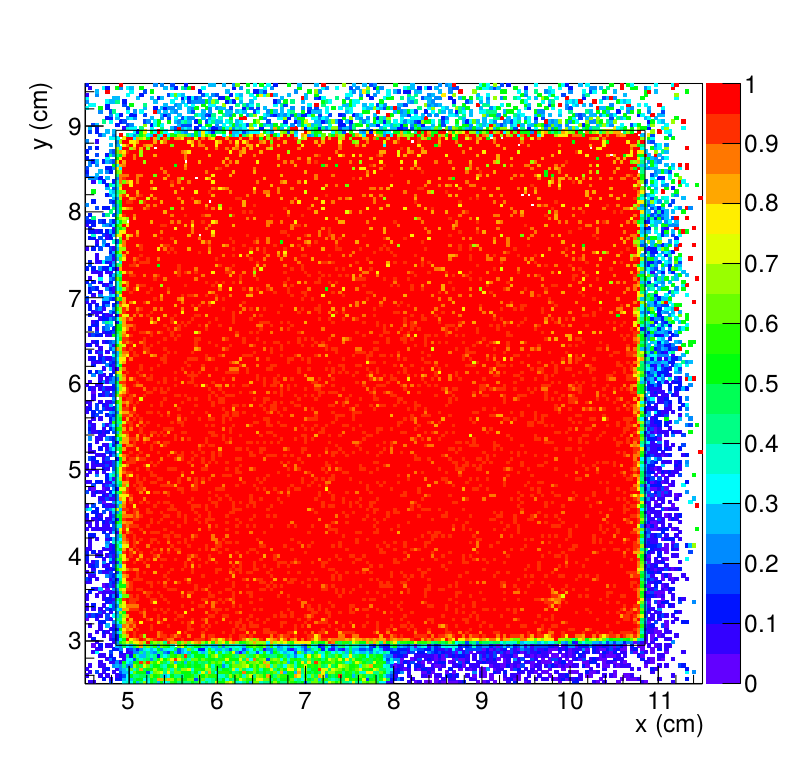}%
	\includegraphics[scale=0.35,type=png,ext=.png,read=.png]{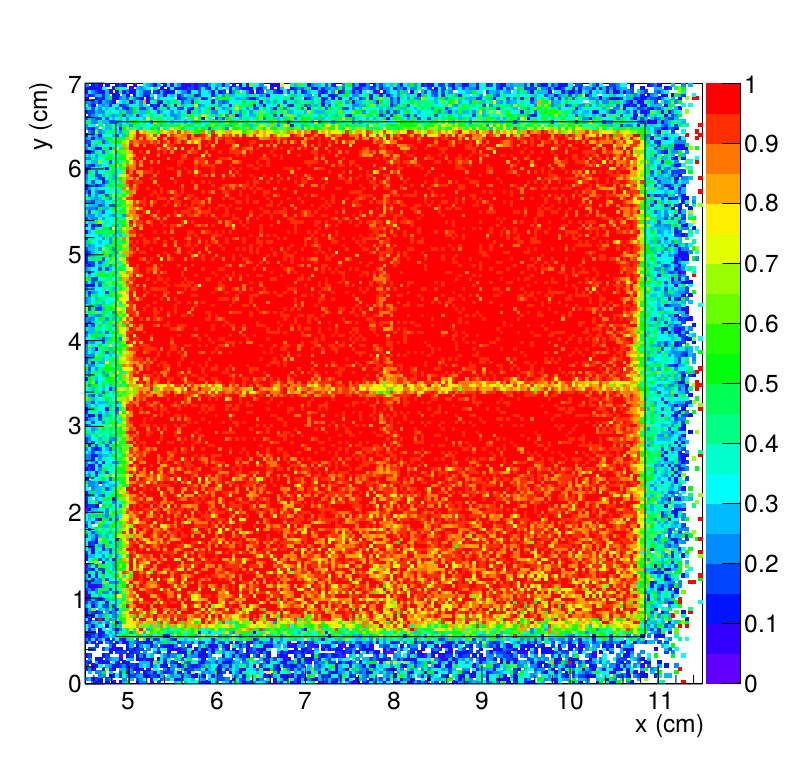}
	\caption{Efficiency map for the PLAS (left) and POLY (right) upstream UCM with a 20 ADC counts threshold (color version online).}
	\label{fig:effmapWG1}
\end{figure}

Efficiency maps are obtained by taking the ratio between the $x$-$y$
2D distribution of the impact point on the upstream face of the
calorimeter for events with a pulse height\footnote{Taken as the sum
  over the four upstream UCMs.} above a certain threshold
(``efficient response'') and the same distribution for all events. No
particle identification information was used. The SiPM were biased at 36~V for both calorimeters. Results are shown in
Fig.~\ref{fig:effmapWG1} for a threshold of 20 ADC counts\footnote{Corresponding to $\sim$~$1/15$ and to $\sim$~$2/75$ of the mip signal in POLY and in PLAS respectively (see Fig.~\ref{fig:WGepi2-3_36bis}).} for
PLAS~(left) and POLY~(right). The pattern of the $6\times 6~$cm$^2$
front face of the calorimeters is clearly visible. Points with lower
efficiency located outside of the calorimeter face (black square)
originate from events occurring in the dead material and contributing
with some energy release in the calorimeter. For POLY the effect is
larger due to the presence of the aluminum container while PLAS was
just wrapped in black tape.  The pattern at low $x$ and low $y$ for
PLAS is given by the presence of another smaller calorimeter which was
positioned in contact with PLAS. The central white plastic cross
used to create the four optically independent compartments in the
polysiloxane of POLY is clearly visible given its thickness of
about 1~mm. In the case of PLAS the four tiles were divided by a
smaller gap of reflective Mylar of about 100~$\mu$m only.

\begin{figure}
	\centering
	\includegraphics[scale=0.14,type=png,ext=.png,read=.png]{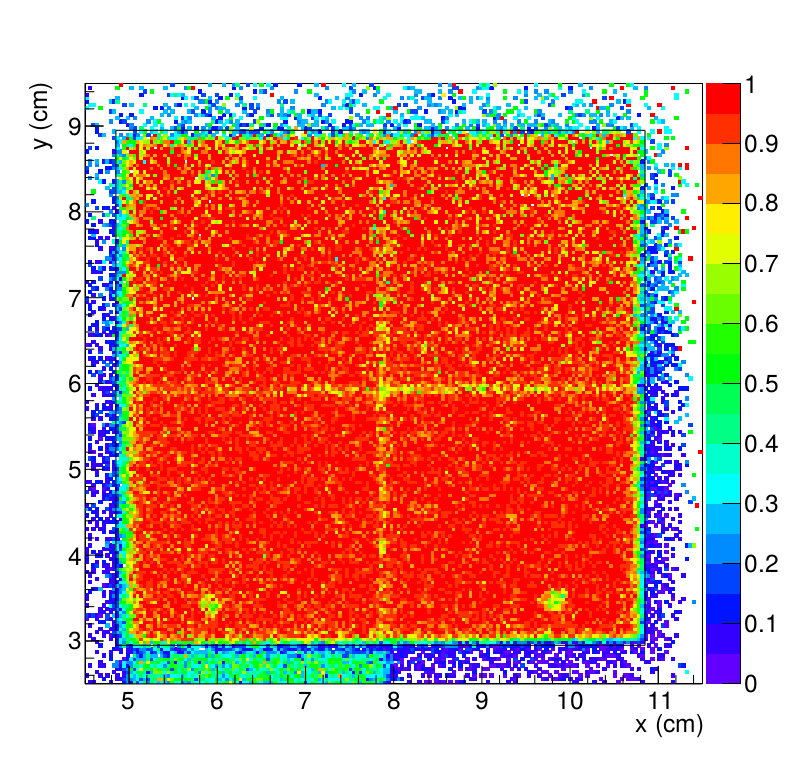}%
	\includegraphics[scale=0.14,type=png,ext=.png,read=.png]{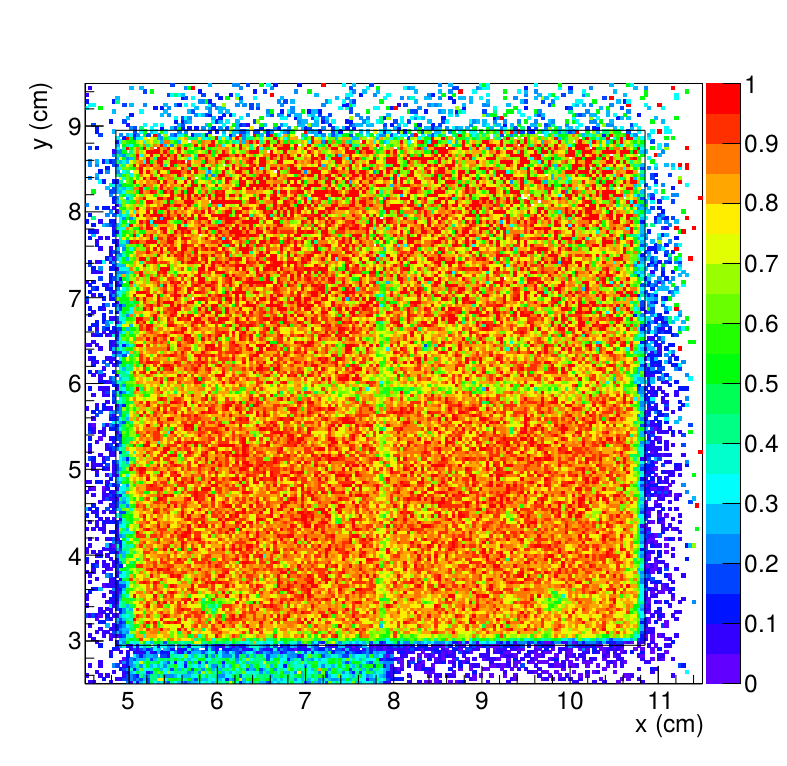}%
	\includegraphics[scale=0.14,type=png,ext=.png,read=.png]{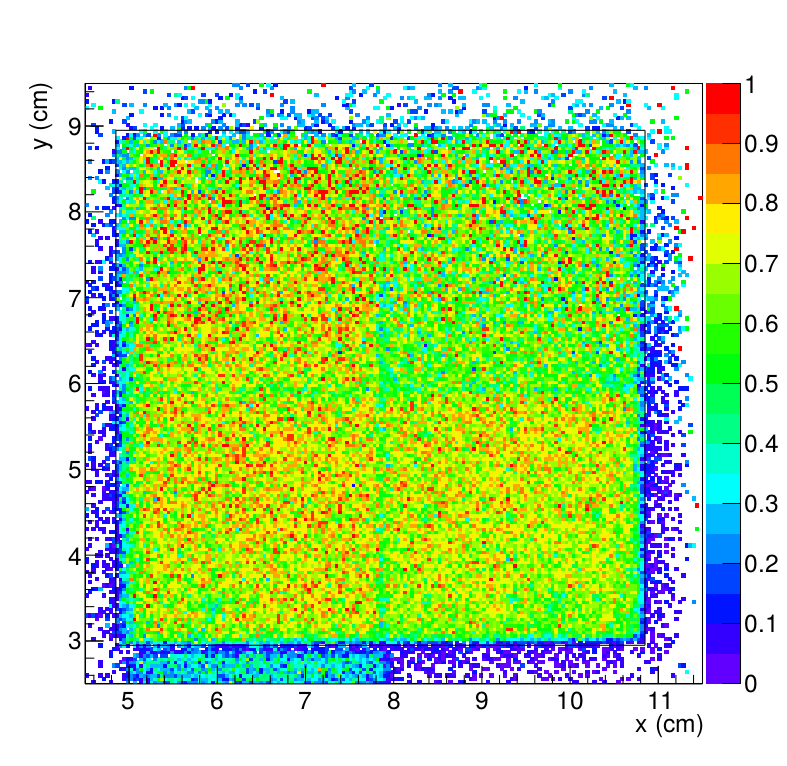}%
	\includegraphics[scale=0.14,type=png,ext=.png,read=.png]{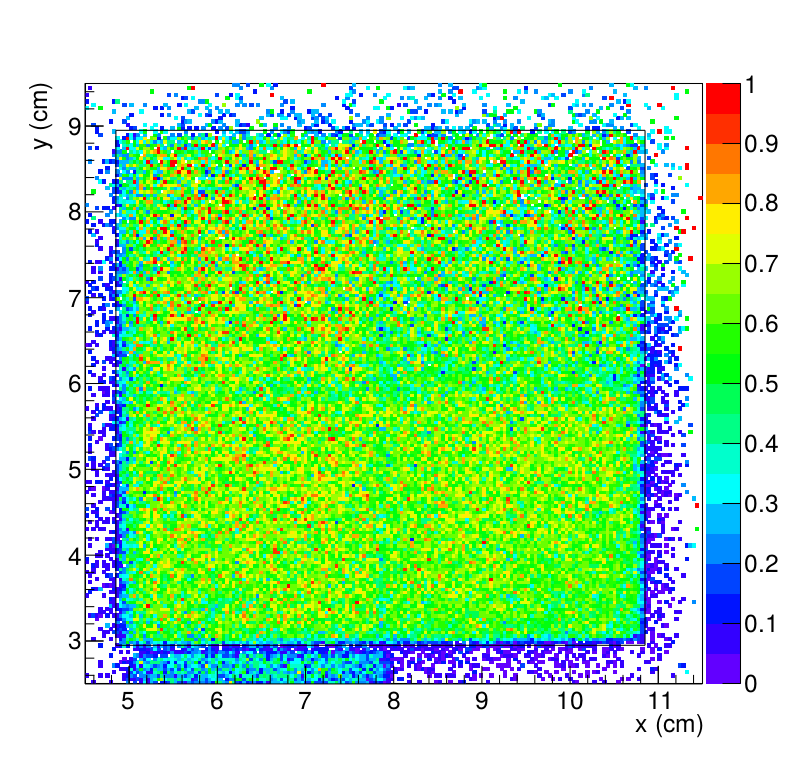}%
	\includegraphics[scale=0.14,type=png,ext=.png,read=.png]{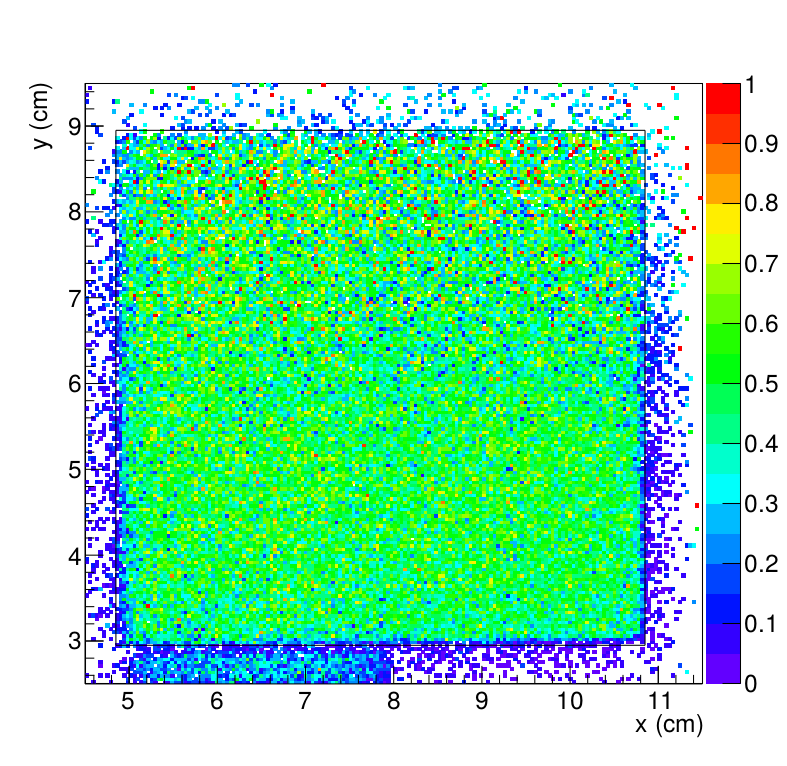}\\
	\includegraphics[scale=0.14,type=png,ext=.png,read=.png]{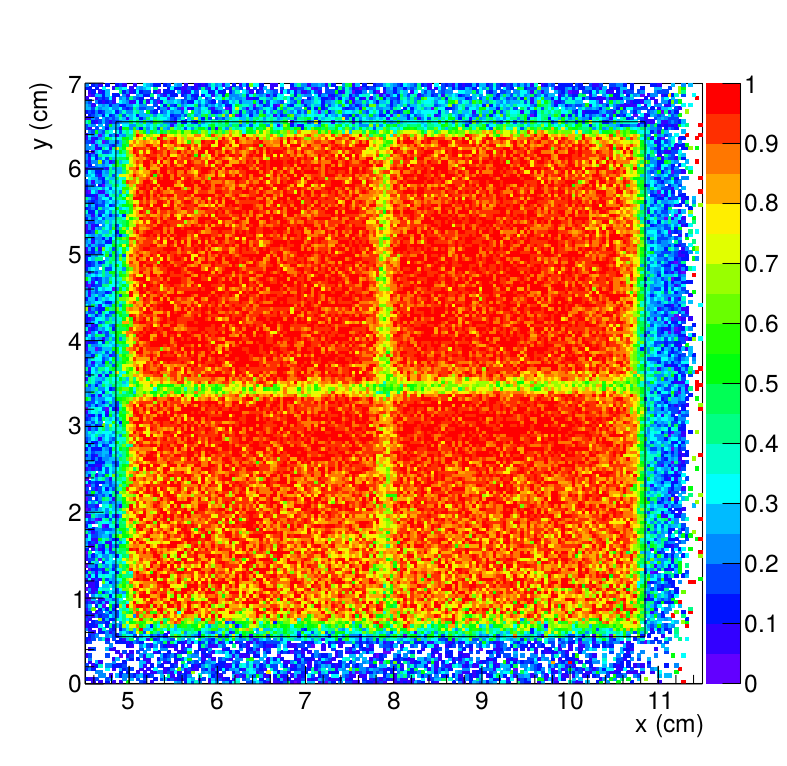}%
	\includegraphics[scale=0.14,type=png,ext=.png,read=.png]{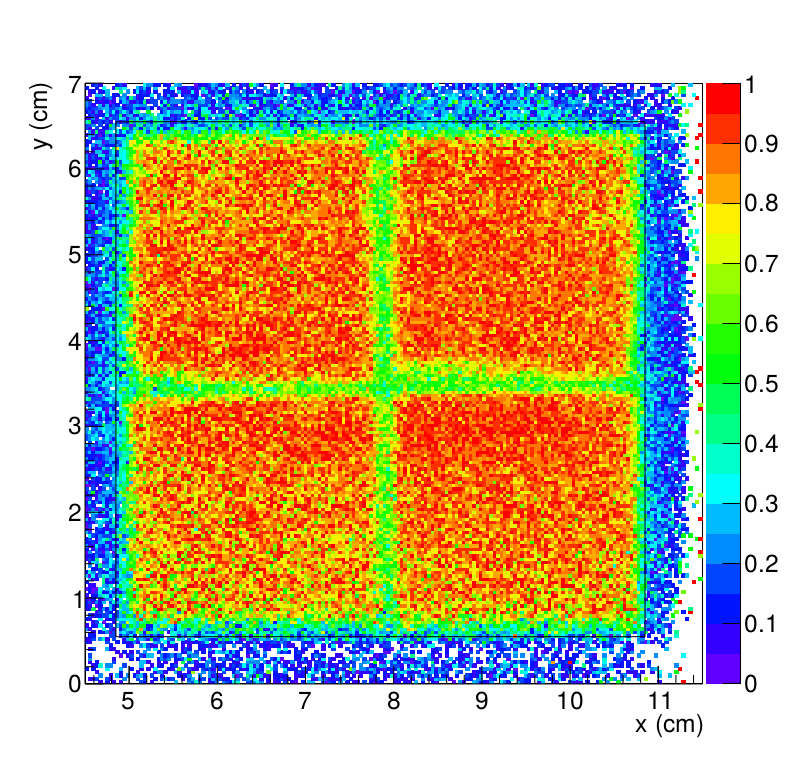}%
	\includegraphics[scale=0.14,type=png,ext=.png,read=.png]{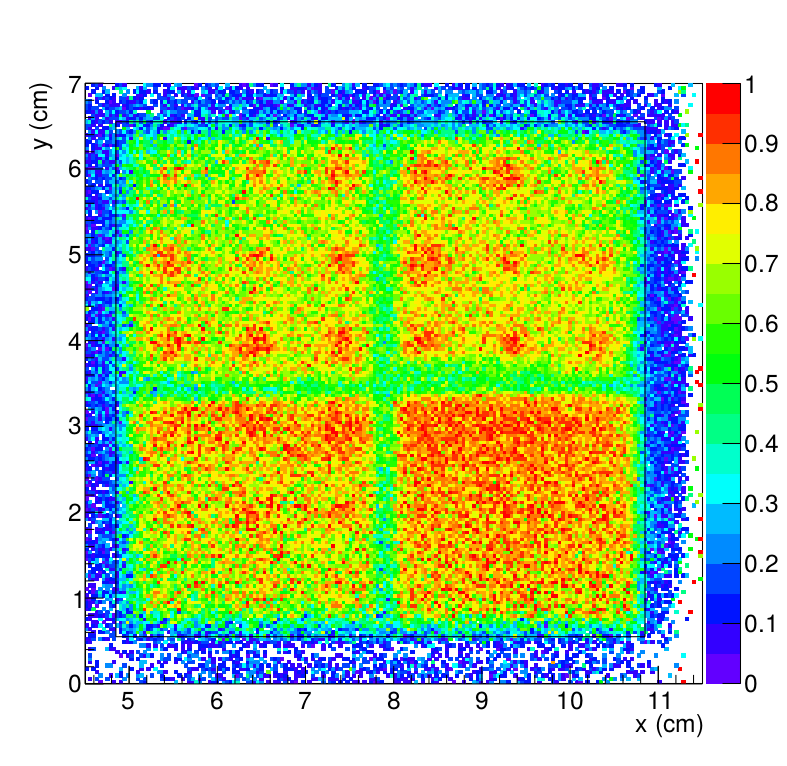}%
	\includegraphics[scale=0.14,type=png,ext=.png,read=.png]{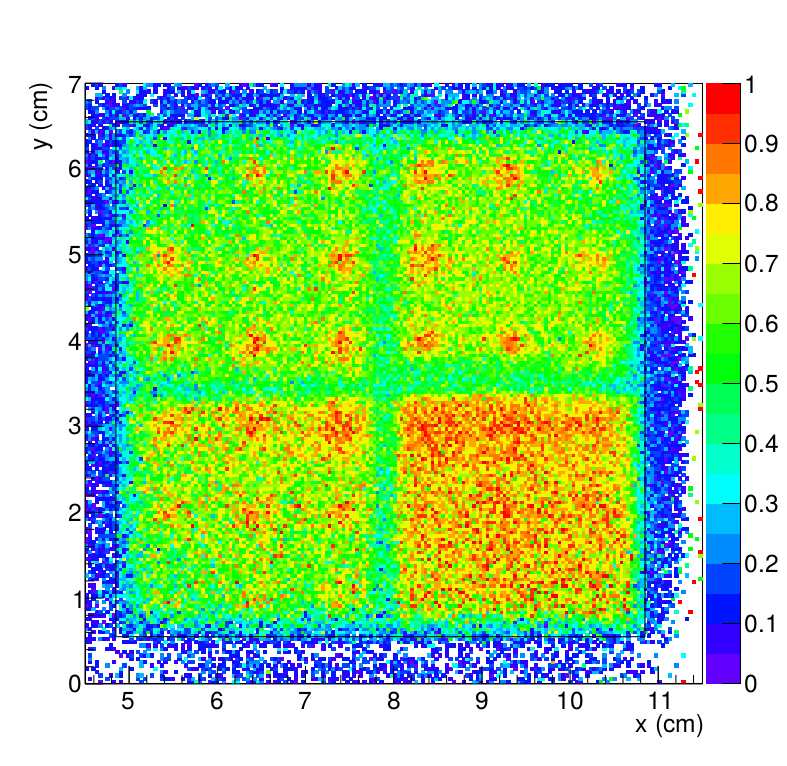}%
	\includegraphics[scale=0.14,type=png,ext=.png,read=.png]{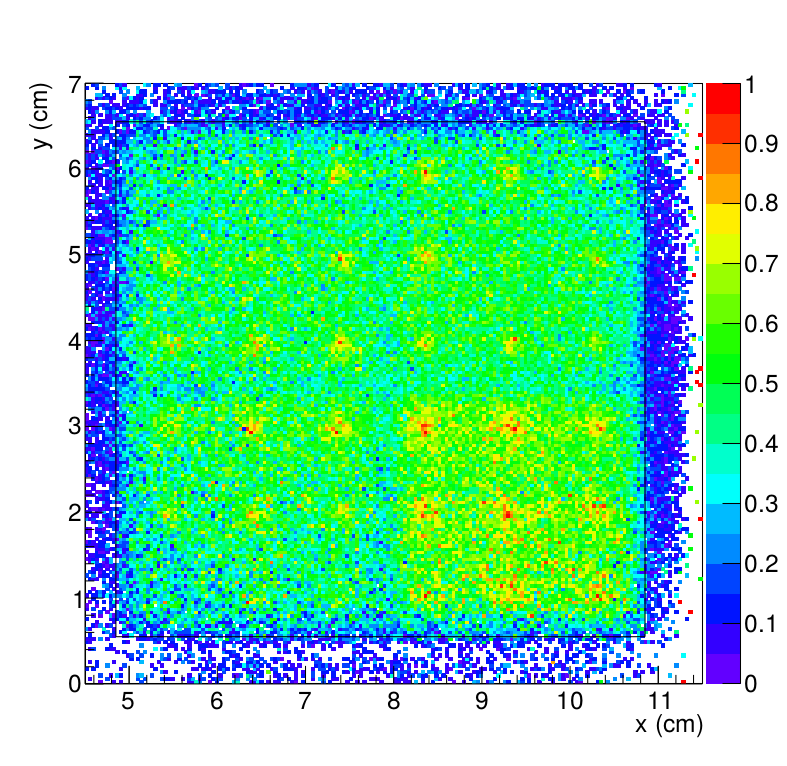}
	\caption{Efficiency maps for increasing thresholds from left
          to right for the most upstream module using all particle
          types. The upper row refers to PLAS the lower one to POLY (color version online).}
	\label{fig:polys1}
\end{figure}

The maps were also
calculated for increasing values of the threshold to seek for smaller dead zones, as shown in
Fig.~\ref{fig:polys1}. Thresholds are $\{$40, 60, 80, 90, 120$\}$ and
$\{$80, 120, 160, 180, 240$\}$ ADC counts for POLY and PLAS
respectively. For PLAS the small gap between the tiles is visible for threshold values above 80. A drop of efficiency is also visible for
thresholds 80 and 120 (above these values particle statistics is too low) in the
positions where the WLS fibers are located ($3\times3$ per tile with a 1~cm spacing). Four larger areas with lower efficiency are also
visible due to the presence of the long bolts spanning the full module
length (visible in Fig.~\ref{fig:FF} right), which were used to tighten the absorber--scintillator stack.
In general the decrease in efficiency has a similar trend for POLY and
PLAS but it can be noticed that POLY retains high efficiency if the
particle passes close to the fiber (pattern of red regions for
thresholds 80, 90, 120) while the decrease is more uniform in space for PLAS. 
This effect is consistent with the results of the optical simulation (see right plot of Fig.~\ref{fig:optisim}), which indicates larger photon collection efficiency for POLY especially in that region.

For POLY in the map corresponding to threshold 80 (third plot from
left in the bottom row) a region with reduced efficiency is observed
that could be due to the presence of bubbles or to the discontinuities in the polysiloxane creating total internal
reflection that were observed visually (see Fig.~\ref{fig:WGpics} left).

It can also be observed 
that for POLY the UCM in the
right-bottom position is on average more efficient, 
most likely due to the better fiber-SiPM mechanical match in that area.

\begin{figure}
\centering
\includegraphics[scale=0.5,type=pdf,ext=.pdf,read=.pdf]{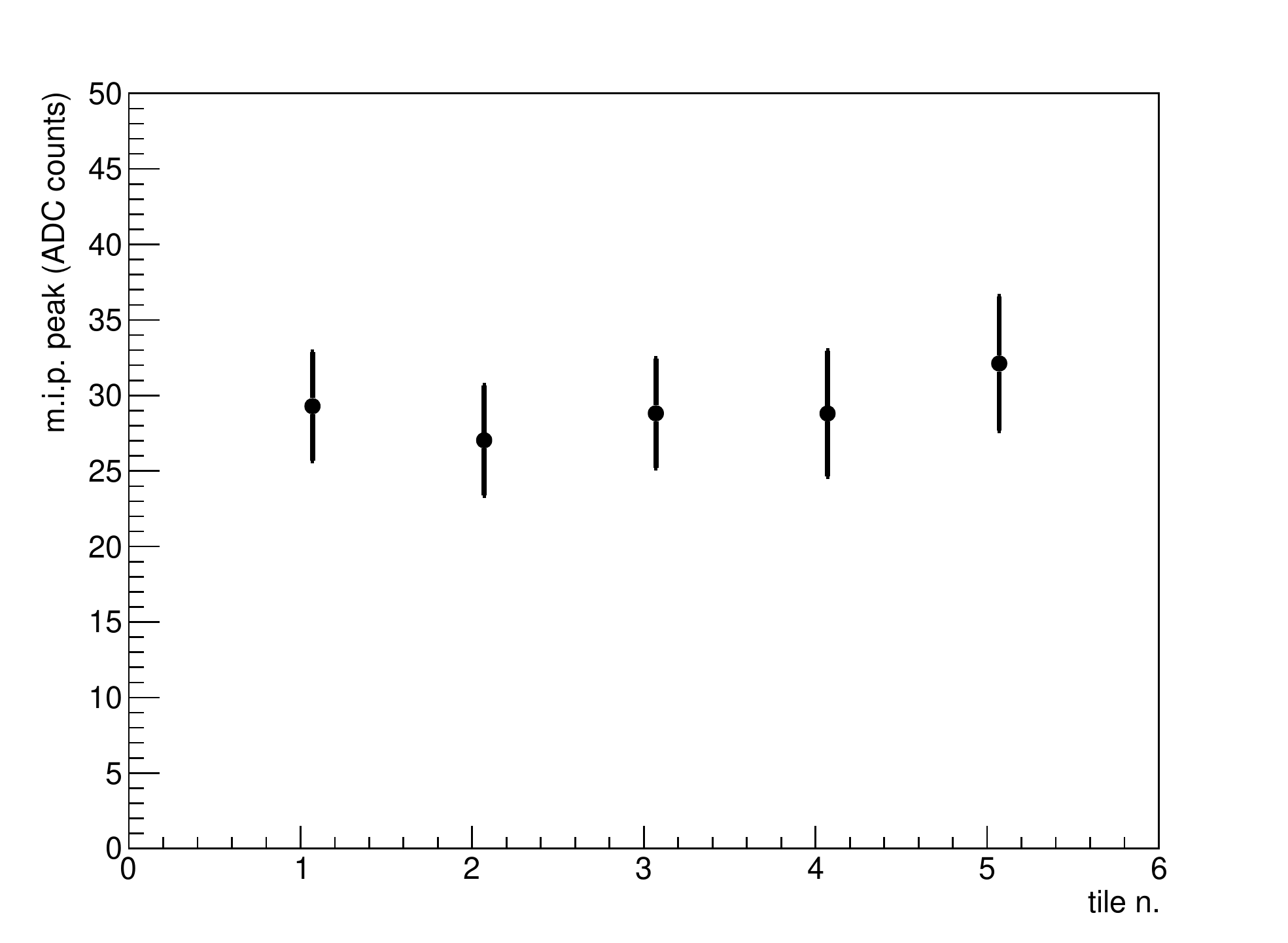}
\caption{Light yield of the scintillator tiles. 
Tiles are ordered from 1 to 5 according to their position along the beam direction.
Markers represent the means of the most probable
  values for Landau fits of mip tracks for each tile position in the UCMs, while their spreads is shown by the error bar.}
\label{fig:effmapWG2}
\end{figure}

The polysiloxane calorimeter was also exposed with a 90$^\circ$ tilt to study the
light yield of individual tiles with mip tracks. We extracted the most probable value parameter of a Landau function model used to fit the signal distribution on each of the 60 (5$\times$4$\times$3) tiles. Results are shown in Fig.~\ref{fig:effmapWG2}. 
%Each colored point corresponds to the most probable value of Landau fits for each tile. 
Tiles are ordered from 1 to 5 according to their position along the beam direction. The mean of the most probable values for each tile position
is given by the marker while their
spread (ranging from 12 to 15 \% of the central value)
is shown by the error bar.

\begin{figure}
\centering
\includegraphics[scale=0.23,type=png,ext=.png,read=.png]{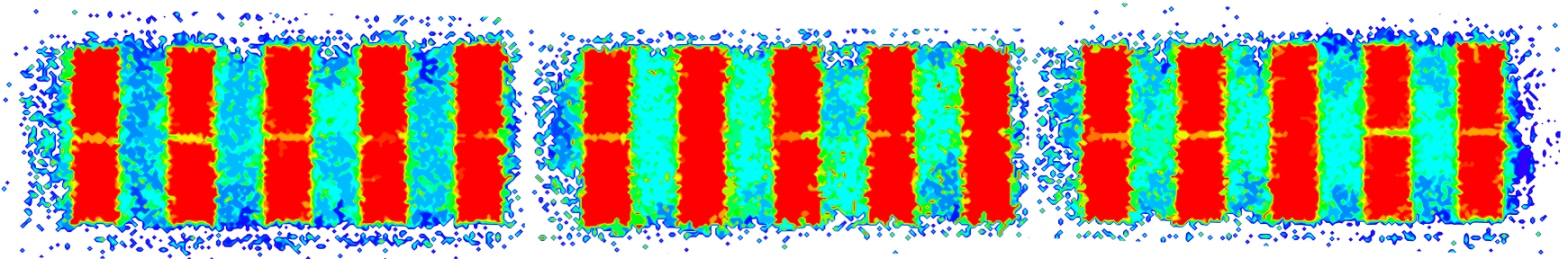}
\caption{Lateral efficiency maps with a 10 ADC threshold for the three
  modules. The free surface of the liquid polysiloxane is down in this
  figure (color version online).}
\label{fig:lat0}
\end{figure}

The efficiency maps of the three modules are shown in
Fig.~\ref{fig:lat0} for a 10 ADC counts threshold. The meniscus effect
bending the free surface of the liquid during preparation is visible
especially on the left bottom.

\begin{figure}
\centering
\includegraphics[scale=0.22,type=png,ext=.png,read=.png]{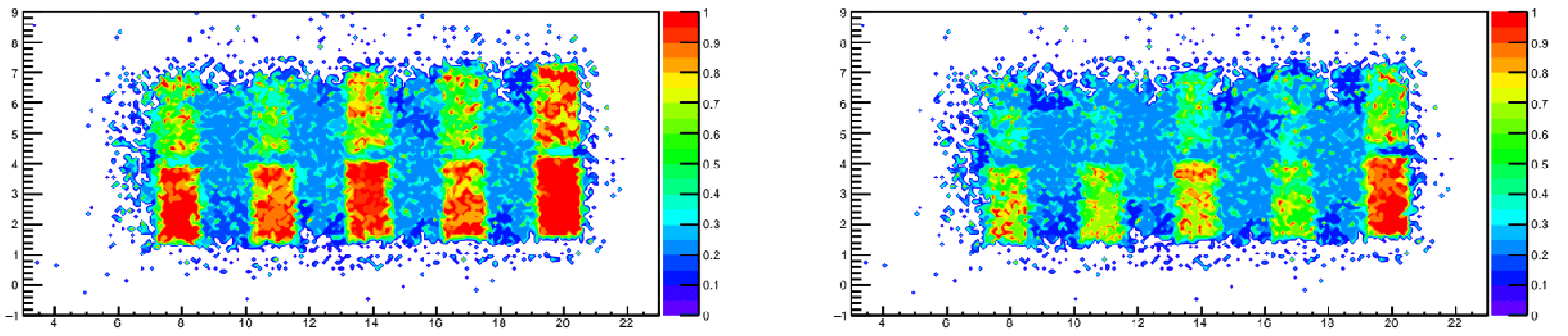}
\caption{Lateral efficiency maps at different thresholds of 30 (left)
  and 35 (right) ADC counts for the upstream module as an example (color version online).}
\label{fig:lat1}
\end{figure}

Figure \ref{fig:lat1} shows the efficiency of the upstream module with
higher thresholds of 30 (left) and 35 (right) ADC counts to enhance the
visibility of non-uniformities. The other two modules show similar
patterns. Again the higher efficiency close to the WLS fibers is
confirmed. In general, the pouring process does not induce any efficiency gradient in the horizontal or vertical direction. Still non-uniformity along the cells are visible and have been traced to the presence of bubbles and on areas where the Tyvek was detached by stresses during the cool down in the production phase.

\subsection{Measurement of light collection efficiency}
\label{ly}
\begin{figure}
\centering
\includegraphics[scale=0.2,type=png,ext=.png,read=.png]{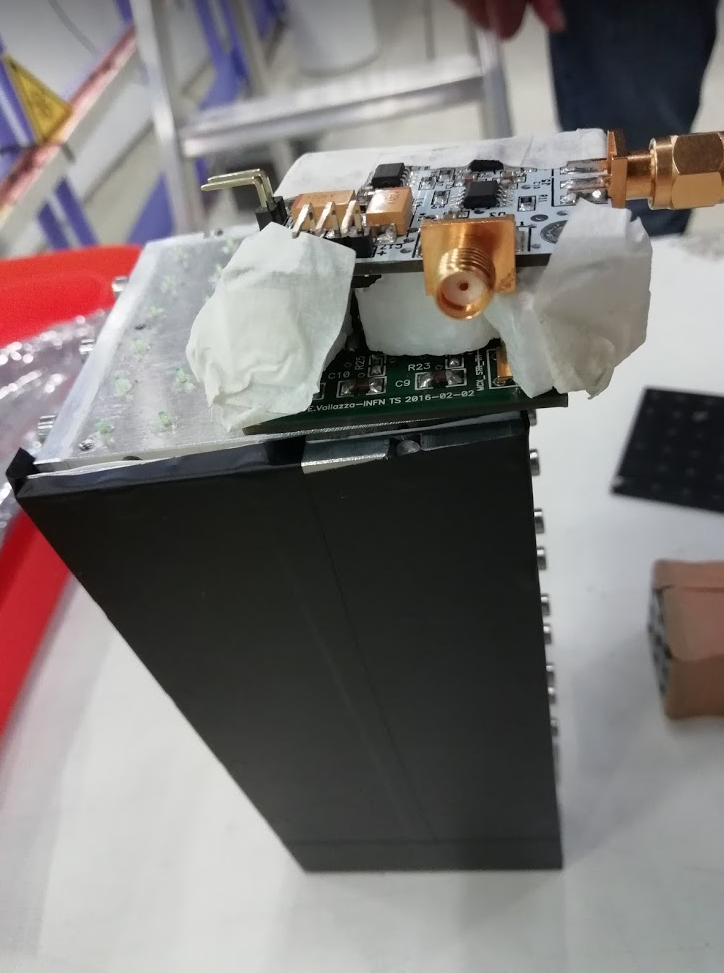}%
{~~~~~~~~}%
\includegraphics[scale=0.39,type=pdf,ext=.pdf,read=.pdf]{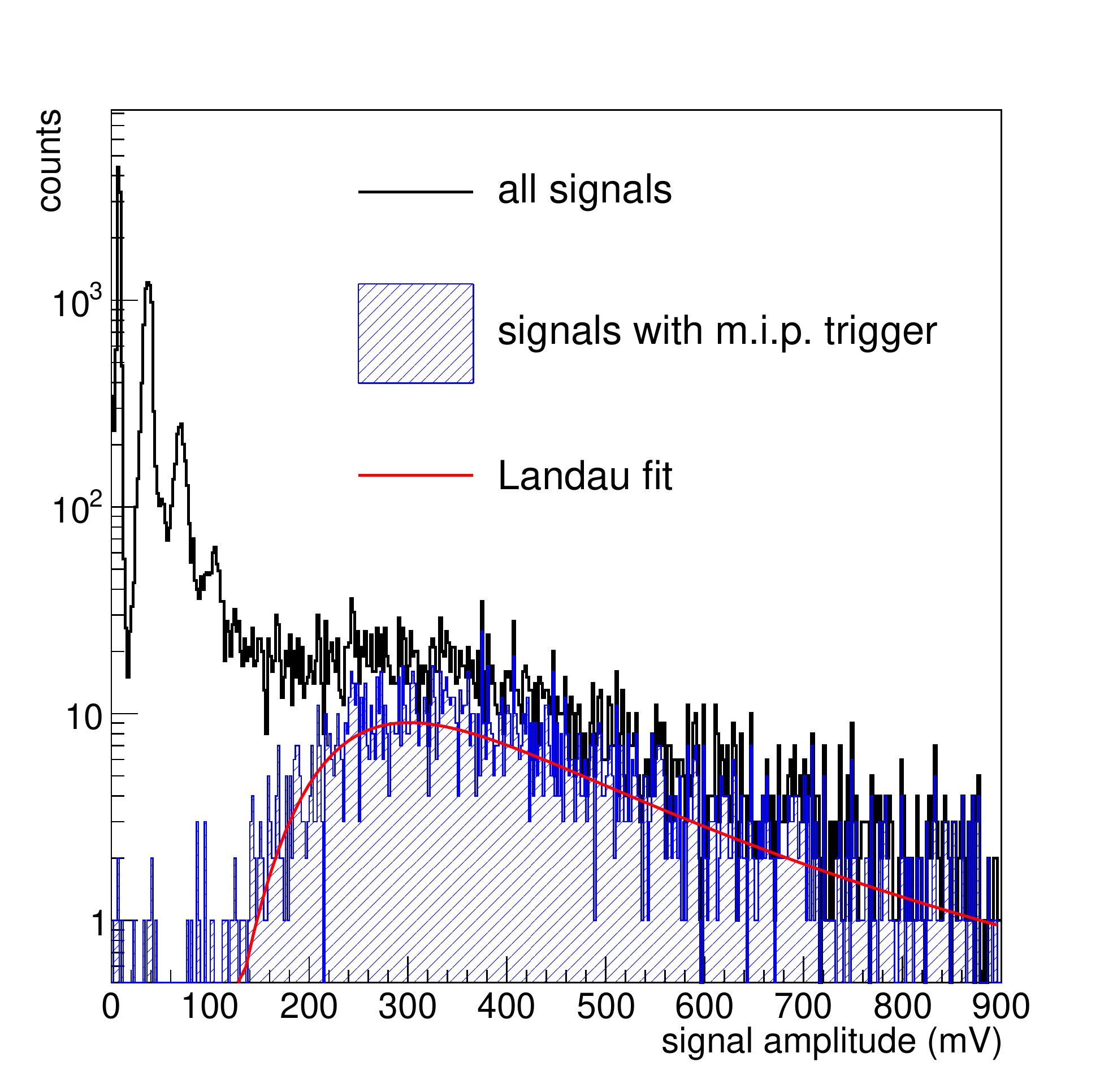}
\caption{Left: the amplifier connected directly on the SiPM PCB for
  the measurement of single photo-electron peaks of the central SiPM
  of the UCM. Right: the spectra of the digitized
  signal amplitude in the acquisition time window for POLY at a
  bias voltage of 38~V. The black histogram shows the distribution for
  all acquired waveforms while the blue filled one (with a Landau function
  fit superimposed) is for the subsample of single-cluster events in
  which the particle enters the fiducial volume of the calorimeter.} 
\label{fig:PE}
\end{figure}

In May 2018 one of the polysiloxane UCM\footnote{The bottom right one
  in Fig.~\ref{fig:polys1} (bottom row) which is also the most
  efficient. NB: the comparison between POLY and PLAS of
  Fig.~\ref{fig:WGepi2-3_36bis} is an average on four UCM (fiducial
  volume across the four UCM).}  was tested after having polished the
WLS fibers at the same level of the aluminum backplane of the
calorimeter module. From now on we will refer to this improved module as POLY$^\prime$. The signal from the central SiPM 
was amplified with a trans-impedance amplifier (model
ASD-EP-EB-N from Advandsid ~\cite{TIA}). The amplifier was connected
directly on an Advansid SiPM with 15 $\mu$m cell size mounted on
a PCB (as shown in Fig.~\ref{fig:PE} left). The low-gain output
providing a 5$\times$ amplification was digitized with a v1751 CAEN
digitizer (10 bit, 1~V range, 1~GS/s). 

The number of collected photoelectrons for a mip was estimated by comparing the signal distribution with the distribution from single photo-electrons (P.E.) produced by noise. The crosstalk effect for the SiPM used was $\sim$20\%, as estimated from the ratio of the counts in the second and in the first P.E. peaks. In the right-most plot of Fig.~\ref{fig:PE} we show the spectra of the maximum
digitized signal amplitude of the UCM in the acquisition time window at a bias voltage of 38~V. The black histogram shows the
distribution for all acquired waveforms while the blue filled one (with a
Landau function fit superimposed) is for the subsample of
single-cluster events in which the particle enters the fiducial volume (2$\times$2~cm$^2$) of the UCM.

\begin{figure}
\centering
\includegraphics[scale=0.73,type=pdf,ext=.pdf,read=.pdf]{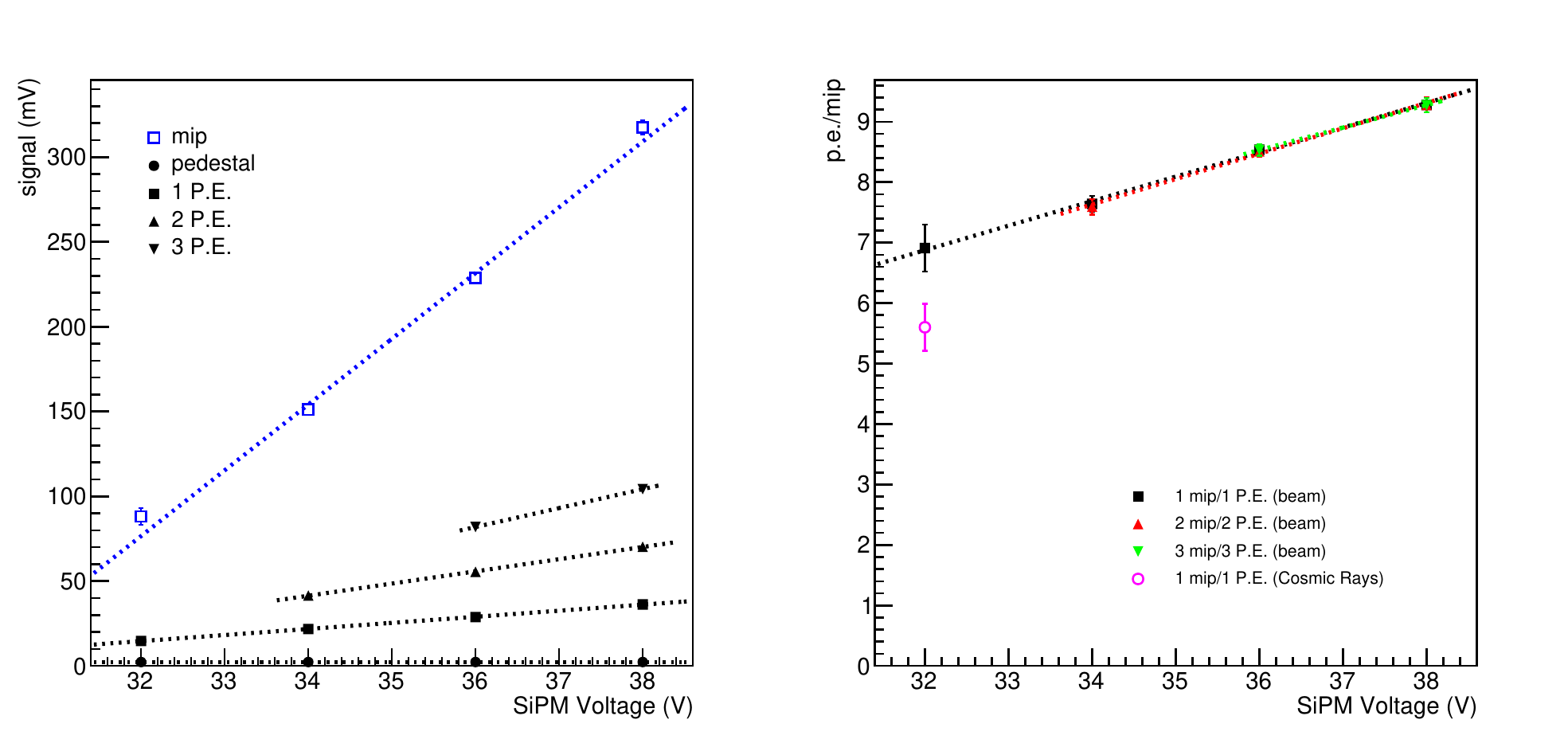}
\caption{Left: Dependence of the position of 1, 2, 3 P.E.  signals
  (black) and of the mip peak (blue) in ADC counts with the applied
  SiPM bias. Right: P.E. per mip as a function of the SiPM bias. The
  measurement obtained with cosmic ray tracks is also reported in magenta (see text for details).}
\label{fig:PE2}
\end{figure}

The dependence of the position of 1, 2, 3 P.E. signals (black) and
of the mip peak (blue) in ADC counts with the applied SiPM bias is
shown in Fig.~\ref{fig:PE2} left. The number of P.E./mip is obtained
as the ratio between the mip peak and the single P.E peak after pedestal subtraction. The same quantity was also estimated using 2 and 3 P.E. peaks when properly visible (higher bias voltage). As expected this ratio has some dependence on the bias voltage since the photon detection
efficiency varies with the applied over-voltage. This is
shown in Fig.~\ref{fig:PE2} right.
At a voltage bias of 36~V we observe about 8.5 P.E./mip corresponding to about 70-80 P.E./mip when reading all the 9 SiPM. The correction is obtained with a Monte Carlo simulation since the light is not exactly equally shared among the nine fibers.

The test was repeated using a cosmic muon sample and
a bias voltage of 32~V for both calorimeters (POLY$^{\prime}$ and PLAS) giving $(5.6 \pm 0.4)$~P.E./mip for POLY$^\prime$ and $(8.3 \pm 0.7)$~P.E./mip for PLAS. The result for POLY$^\prime$ is consistent with the measurement done at CERN though slightly smaller (magenta point at 32~V in Fig.~\ref{fig:PE2} right). After normalizing to the same scintillator thickness (10~mm for POLY vs 15~mm for PLAS), the 40\% better collection efficiency ($\epsilon_{coll}$) expected for POLY and the reduced light yield (measurement with sources) we obtain that POLY$^{\prime,corr} = \rm{POLY}^{\prime}\times 2.4 /(1.5 \times 1.4) = (6.4~\pm~0.5)$~P.E./mip. This can be then compared to the $(8.3 \pm 0.7)$~P.E./mip found for PLAS. The light yield of POLY$^{\prime}$ is therefore 30\% smaller than expected. The reduction
in the discrepancy (from a factor 2, as estimated in Sec.~\ref{pid} for the first version of POLY, to 1.3) can be attributed to the improvements in the
optical match between the WLS fibers and the SiPM, after the aforementioned fiber polishing procedure (i.e. reduced lateral
and longitudinal displacements and improved polishing of
the WLS).

\section{Conclusions}
\label{conclusions}

We have developed a 13 $X_0$ shashlik calorimeter using for the first
time a polysiloxane-based scintillator with 1.5~cm thick iron absorbers and
1.5 cm thick scintillator tiles read out by Y11 multi-clad WLS
fibers. The detector was tested with particle beams in October 2017
and May 2018 and compared to a similar calorimeter composed of
standard plastic scintillator (EJ-204, 1 cm thick, with BCF92 WLS
fibers).

The calorimeter provides particle identification capabilities and
energy resolution for electrons at the same level of those obtained
with standard plastic scintillators. 

The prototype demonstrates that the concept of using polysiloxane for
shashlik calorimeters is a viable option with the advantage of reduced
efforts for machining a large number of holes and a much higher
radiation hardness. The fact that the WLS fibers have no air interface
with the scintillator does not compromise the light trapping mechanism and transport to SiPM.
This prototype also proved that is possible to pour the scintillator
through optically separated compartments without significant deterioration of the light yield due to
formation of bubbles or empty volumes near the surfaces. Further improvements
in the uniformity of the tiles (presently at the 15 \% level) and in the coupling between the SiPM and fibers can be envisaged ameliorating the gluing of the Tyvek and the mechanics of the SiPM holder.

\section*{Acknowledgements}

The project leading to this application has received funding from the
European Research Council (ERC) under the European Union's Horizon
2020 research and innovation programme (grant agreement N. 681647).
We thank L. Gatignon, M. Jeckel and H. Wilkens for help and
suggestions during the data taking on the PS-T9 beamline. We are
grateful to the INFN workshops of Milano Bicocca and Padova for the
construction of the detector and, in particular, to L. Ramina.

\end{document}